\documentstyle[12pt]{report}
\begin{document}
\def\be{\begin{equation}}
\def\ee{\end{equation}}
\def\bearst{\begin{eqnarray*}}
\def\eearst{\end{eqnarray*}}
\def\peleven{\parbox{11cm}}
\def\peffec{\peight{\bearst\eearst}\hfill\peleven}
\def\pspace{\peight{\bearst\eearst}\hfill}
\def\ptwelve{\parbox{12cm}}
\def\peight{\parbox{8mm}}
\def\bear{\begin{eqnarray}}
\def\eear{\end{eqnarray}}
\def\E{{\rm e}}
\input epsf.tex
\newcommand{\slp}{\raise.15ex\hbox{$/$}\kern-.57em\hbox{$\partial$}}
\newcommand{\sla}{\raise.15ex\hbox{$/$}\kern-.57em\hbox{$a$}}
\newcommand{\slA}{\raise.15ex\hbox{$/$}\kern-.57em\hbox{$A$}}
\newcommand{\slB}{\raise.15ex\hbox{$/$}\kern-.57em\hbox{$B$}}
\newcommand{\slD}{\raise.15ex\hbox{$/$}\kern-.57em\hbox{$D$}}
\newcommand{\slb}{\raise.15ex\hbox{$/$}\kern-.57em\hbox{$b$}}
\newcommand{\slW}{\raise.15ex\hbox{$/$}\kern-.57em\hbox{$W$}}
\font\grrm=cmbx10 scaled 1200
\font\vb=cmbx10 scaled 1440
\font\bigcal=cmsy10 scaled 1200
\def\eightpoint{\def\rm{\fam0\eightrm}}
\def\flex{\raise 6pt\hbox{$\leftrightarrow $}\! \! \! \! \! \! }
\def\tr{ \mathop{\rm tr}}
\def\atanh{\mathop{\rm atanh}}
\def\Tr{\mathop{\rm Tr}}
\def\dal{\Box} 
\def\Natural{\hbox{\hskip 1.5pt\hbox to 0pt{\hskip -2pt I\hss}N}}
\def\Integer{\>\hbox{{\sf Z}} \hskip -0.82em \hbox{{\sf Z}}\,}
\def\Rational{\hbox{\hbox to 0pt{\hskip 2.7pt \vrule height 6.5pt
                                  depth -0.2pt width 0.8pt \hss}Q}}
\def\Real{\hbox{\hskip 1.5pt\hbox to 0pt{\hskip -2pt I\hss}R}}
\def\Complex{\hbox{\hbox to 0pt{\hskip 2.7pt \vrule height 6.5pt
                                  depth -0.2pt width 0.8pt \hss}C}}
\def \ln {{\rm ln}\, }
\def \cotg {\rm cotg }

\vskip 1cm
\begin{tabbing}
\hskip 11.5 cm \=  \\
\>January 1997
\end{tabbing}
\vskip 1cm
\begin{center}
{\Large\bf Two-dimensional Quantum Field Theory, examples and applications}
\vskip 1.1cm
{\large \bf E. Abdalla$^{a}$}\\
Instituto de  F\'\i sica-USP, C.P. 66.318, S. Paulo, Brazil\\
and\\
ICTP, High-Energy Division \\
\vskip 0.3cm
eabdalla@snfma1.if.usp.br
\end{center}
\abstract 

\noindent

The main principles of two-dimensional quantum field theories, in 
particular two-dimensional QCD and gravity are reviewed. We study 
non-perturbative aspects of these theories which make them particularly 
valuable for testing ideas of four-dimensional quantum field theory. 
The dynamics of confinement and theta vacuum are explained
by using the non-perturbative 
methods developed in two dimensions. We describe in detail how the 
effective action of string theory in non-critical dimensions can be 
represented by Liouville gravity. By comparing the helicity 
amplitudes in four-dimensional QCD to those of integrable self-dual 
Yang-Mills theory, we extract a four dimensional version of
two dimensional integrability.

\vfill\eject

\tableofcontents
\chapter{General setup}
As a natural extension of Quantum Mechanics, Relativistic Quantum Field
Theory has demonstrated its predictive power in the 
calculation of processes in Quantum Electrodynamics. There are however 
conceptual and technical difficulties, since the local products
of quantum fields, which are operator-valued distributions,
are ill defined. This problem can only be resolved via the techniques of 
renormalization.

The general non-perturbative properties of quantum field theory were first
extracted from a perturbative setup by the so-called $LSZ$ formalism. 
Next, dispersion relations were found and were
used to obtain non-perturbative information. These developments were 
followed by the axiomatic approach, known as constructive $QFT$, where 
functional analysis is the main technique. An important consequence of
this approach was the PCT theorem connecting spin and statistics.

Meanwhile, all dynamical calculations in $QFT$ were restricted to perturbation
theory. This rendered calculations involving strong interactions unreliable.
In addition information about the bound state spectrum could only  be accessed
within approximate non-perturbative - and often  non-unitary--schemes.
As a result, $QFT$ fell into stagnation and was discredited for many years.
These difficulties  provided, in particular, a motivation for
$S$-matrix theory. The predictive power of this
theory turned out to be very small, since it was entirely based on kinematical
principles, analyticity and the bootstrap idea.  An underlying
dynamical framework was lacking. Nevertheless, analyticity in the complex
angular momentum plane led to the important concept of duality.
An explicit realization of these concepts by the remarkable
Veneziano formula led to a new parallel development, in the sixties,
the dual models. However, the predictions of the dual models
at high-energy scatterings were incorrect. Moreover, an analysis of the
pole structure of higher-order corrections required the introduction of a
somewhat mysterious new concept, the pomeron.

In the meantime, $QFT$ had scored remarkable successes in the realm of the
weak interactions. Moreover, symmetry principles had proven powerful in 
predicting the masses of strongly interacting particles 
without the recourse to dynamical calculations.
This led to a revival of $QFT$ in the late sixties. Consequently, in the
seventies, much attention was given to the non-perturbative aspects.
Quantum chromodynamics ($QCD$) was proposed as the fundamental theory of
the strong interactions. However, reliable non-perturbative calculations were 
still lacking in four dimensions and were only available for specific models
in two-dimensional space-time\cite{aar}. It was understood that the short 
distance singularities of quantum field theory play
a key role in the dynamical structure of the theory\cite{31}. The experimental
results on lepton-proton scattering at large momentum transfer, required that a
realistic theory of the strong interactions be asymptotically free.

The first soluble model, describing the current-current
interaction of massless fermions was discussed by Thirring\cite{thirring}
in 1958, as an example of a completely soluble quantum field theoretic model 
obeying the general principles of a $QFT$\cite{klaiber}.
Subsequently, Schwinger\cite{schwinger} obtained an exact solution 
of Quantum Electrodynamics in 1+1 dimensions, $QED_2$. A number of interesting
properties, such as the nontrivial vacuum structure of this model, were however
only later revealed\cite{loswie}, and it was found that there is a long range 
Coulomb force for the charge sectors of the theory. This long range force was 
interpreted as being responsible for the confinement of quarks\cite{cakosuss}. 
The problem of confinement and the related phenomenon of screening of charge 
quantum numbers in two dimensions have been extensively 
studied\cite{screening1,screening2}, and have served as a basis for sharpening 
the similar concepts in higher dimensions.
The surprisingly rich structure of two-dimensional quantum electrodynamics was
found to describe several important features of non-abelian gauge theories,
which were under investigation in the seventies. 

Several further developments of increasing importance followed.
Two-dimensional models which were exactly integrable classically were
extensively  studied\cite {33}. Such integrable models are
generally characterized by the existence of an infinite number of conservation
laws\cite{34}. In the cases where these conservation laws survive
quantization, the $S$-matrices and their associated monodromy matrices can
be computed exactly. Some of the results concerning classical 
integrability have also been generalized to higher dimensions\cite {41}.

In the framework of two-dimensional models, the possibility of writing fermions
in terms of bosons (bosonization) has been a powerful method for obtaining
non-perturbative informations. The building blocks of the
bosonization scheme are the exponentials of the free bosonic fields. The
fermion number of this composite operator is directly linked to the infrared
behaviour of the zero mass scalar fields. This leads to a superselection
rule\cite {49}, which makes the charged sectors appear in a very natural way.

The bosonization techniques become cumbersome when applied to
non-abelian theories. Significant progress in the direction of non-abelian
bosonization has been achieved\cite {50,wittencmp}, and
an equivalent bosonic action involving the principal sigma
model plus a Wess-Zumino term was obtained.

Non-linear sigma models have a long history. A particularly important role has
been played by the class of two-dimensional integrable non-linear sigma models,
which have a geometrical origin\cite{eichenherr}. They have been shown to share
several properties with Yang-Mills theories in four
dimensions\cite{eichenherr,dadda}. When quantized,
non-linear sigma models also exhibit features, believed to be properties of
realistic theories, such as dynamical mass generation, and a long range 
force\cite{dadda} for simple gauge groups\cite{abdforgo}, disappearing after a
suitable interaction with fermions (supersymmetric or minimal), which
liberate the partons\cite{75}. These properties
make them appealing as toy models for the strong interactions\cite{57}.
However, their geometrical origin makes them also very interesting mathematical
objects to be studied in their own right, as well as in string theory.

Two-dimensional space-time has also proven to be an excellent laboratory for
the study of gauge-anomalies and the consistency of anomalous chiral gauge
theories. The exact solubility of two-dimensional chiral $QED$\cite {71,73}
has played here an important role in opening up a whole new line of
developments in the area of chiral gauge theories.

More recently, it has been shown that in two-dimensional quantum field 
theories, Poincar\'e and scale invariance imply the existence of an infinite
dimensional symmetry group\cite{bpz}. As a result, non trivial correlators can
be  exactly computed. They are found to be  related to solutions of
hypergeometric differential equations. The parameters labelling these
equations, which are regarded as the critical indices, have been classified and
characterize the correlators uniquely\cite{bpz}. The above ideas may be 
generalized to include the interaction with conformally invariant 
gravity\cite{pol}. In the light-cone-gauge the theory simplifies 
dramatically, due to a new $SL(2,R)$ symmetry\cite{pol,az}. The critical 
indices of the theory may be computed from a very simple equation relating 
them to the critical indices of the theory in flat space. The results have 
also been generalized to the supersymmetric case\cite{pz-az}.

To summarize, two-dimensional models have been an extraordinary laboratory to
test ideas in quantum field theory. The Thirring model provided a
realization of an exactly soluble quantum field theory, while the Schwinger and
the non-linear sigma models were found to exhibit properties  of four
dimensional non-abelian gauge theories. However, two-dimensional $QFT$ also
plays a direct role in the description of physical reality, having applications
in string theories, as well as in statistical mechanics. In particular, the
methods developed in two-dimensional $QFT$ have been used to extract results
concerning the critical behavior of models in statistical mechanics, using
conformal invariance alone\cite{bpz}. An extraordinary amount of physically
interesting\cite{85} as well as mathematically elegant concepts\cite{86} have
emerged from the study of such theories. There is a deep relationship 
between rational conformal invariance in two-dimensional space-time
and the Chern-Simons action in three dimensions\cite{88}. The 
Chern-Simons action is a key element in the generalization of the 
fermion-boson equivalence to three dimensional space-time\cite {90}, as
well as playing an important role in the understanding of 
non-abelian anomalies of chiral gauge theories in any dimension.
Beyond their status as a theoretical laboratory, and their applications in
string theories and statistical mechanics, the study of these models has
led to  recent developments which  opened new possibilities for applications 
of some of the above methods in the study of quantum field theories in higher
dimensions. High-energy scattering amplitudes involving fields with
definite helicity in four-dimensional Quantum Chromodynamics have a rather
simple description, presumably related to integrable models\cite{bardeen},
the same being true for high-energy scattering\cite{fadeevkor}. In
the former case, the scattering amplitudes are related to solutions of
self-dual Yang-Mills equation, while in the latter case the interaction of
external particles is described by the two-dimensional Heisenberg Hamiltonian
of spin systems.
\chapter{A survey of results}
Here we restrict ourselves to the main results obtained in two-dimensional
quantum field theory which had important consequences for realistic models.
In subsequent chapters, the cases of more recent interest are reviewed
in more detail, as well as recent applications to four dimensional quantum
chromodynamics.
\section{Schwinger model}
The Schwinger model, Quantum Electrodynamics in two-dimensions, can be 
exactly solved in terms of the prepotentials for the gauge fields,
\be
A_\mu =-{\sqrt\pi\over e}
\left(\tilde\partial_\mu\Sigma +\partial_\mu\tilde\eta\right)\; .
\label{a-prepot}
\ee
This, together with the bosonisation formulae,
\be
J^\mu\equiv\overline\psi\gamma^\mu\psi=-{1\over\sqrt\pi}\tilde\partial^\mu
\Phi\quad,\quad
J^\mu_5\equiv\overline\psi\gamma^\mu\gamma^5\psi=-{1\over\sqrt\pi}\partial^\mu
\Phi\; ,\label{sch-currents}
\ee
the field equations and the anomaly equation,
\be
\partial_\mu J^\mu_5={e\over 2\pi}\epsilon_{\mu\nu}F^{\mu\nu}\; ,
\label{anoma-sch}
\ee
leads to the conclusions that $\Phi$ and $\Sigma$ differ by a massless
field. That is to say, the current has a transverse part given in terms of 
the (physical)
field $\Sigma$ and a longitudinal part given in terms of the massless fields
(which in turn have to be weakly zero). This condition defines the space of
states. That is,
\be
J^\mu=-{1\over\sqrt\pi}\tilde\partial_\mu\Sigma +L_\mu\label{curr-trans-long}
\ee
where
\be
L_\mu=-{1\over\sqrt\pi}\tilde\partial_\mu \varphi\; .
\ee
The Maxwell equation implies
\be
\tilde\partial_\mu \left( \Box +{e^2\over\pi}\right)
\Sigma -{e^2\over\sqrt\pi} L_\mu =0\; .
\ee
The only physical excitation of the theory is $\Sigma$, which is a free
field of mass ${e^2\over\pi}$, and the physical Hilbert space is characterized
by requiring that the expectation value of $L_\mu$ vanishes. Observables are 
those operators which leave the Hilbert space invariant, that is
\bearst
&&\left\langle\psi'\vert L_\mu\vert\psi\right\rangle =0\\
&&\left.{\cal O}\vert\psi\right\rangle \in{\cal H}_{phys}, \quad {\hbox {if}}
\left.\vert\psi\right\rangle\in{\cal H}_{phys}
\eearst
Consequently, operators such as
\be
\sigma_\alpha=e^{i\sqrt\pi\gamma^5_{\alpha\alpha}(\varphi +\eta)
+ i\sqrt\pi\int_{x^1}^\infty dz^1\partial_0(\varphi +\eta)}
\ee
are observables, but act as constant operators, defining the $\theta$
vacuum structure of the model. The massless Schwinger model
is characterized by the fact that physical fermion fields are not
observables, but flavour quantum numbers are, since the force
between distant quarks does not grow with distance. This fact characterizes
screening. The massive model on the other hand, displays (with the exception
of some $\theta$ worlds) a long-range force, linearly growing with
distance, which means confinement.

\section{Non-abelian generalization}
It is desirable to generalize these results for two-dimensional quantum 
chromodynamics. For non-abelian gauge groups, bosonisation is more
sophisticated: while
in the abelian case fermions can be written in terms of bosons by means
of an exponential map, in the non-abelian case such a map does not exist.

Bosonisation of two-dimensional QCD is achieved by writing the fermionic
determinant in terms of a bosonic integration. Using the 
parametrization $ eA_+=U^{-1}i\partial_+U$, $A_-=Vi\partial_- V^{-1}$,
the effective bosonic action $S_F[A,w]$ for  the fermions is written in terms 
of the so-called gauged Wess-Zumino-Witten action,
\bear
S_F[A,g]=\Gamma[g]+{1\over 4\pi}\!\int\!{\rm d}^2x\,{\rm tr}\,\left[e^2A^\mu A_
\mu-e^2A_+gA_-g^{-1} -eiA_+g\partial_-g^{-1} -eiA_-g^{-1} 
\partial _+g\right]\quad ,\nonumber\\
S_F[A,g]\equiv\Gamma[UgV^{-1}]-\Gamma[UV]\quad .\label{gauged-wzw-action}
\eear
We thereby obtain the $QCD_2$ partition function
\bearst
{\cal Z}=\int{\cal D}A{\cal D}g \E^{i\left( S_{YM} +S_F \lbrack A,g\rbrack 
\right)} \quad .
\eearst

Using algebraic identities, such as the Polyakov-Wiegmann relation
\be
\Gamma [AB]=\Gamma [A]+\Gamma [B]+{1\over
4\pi }\int d^2x {\rm tr} [(A^{-1}\partial _+ A)(B\partial _-
B^{-1})]\quad ,
\ee
we arrive at the partition function for the full QCD$_2$, that is
\be
Z=Z^{(0)}_FZ^{(0)}_{gh} \hat Z_{gh+}^{(0)}
Z_{\tilde W^\prime}Z_{\beta^\prime}\quad ,
\ee
where the first term represents free fermions, the second and third are
the ghost terms, the fourth represents the negative metric excitations.
These terms are all conformally invariant and thus describe the vacuum
excitations. The last term in the above expression describes
the massive excitations. However, all these fields are non-trivially connected
by means of BRST constraints. Some of these constraints are derived from
those already describing the original gauge fields, but there are constraints
appearing from the various changes of variables which one uses to decouple the
theory at the lagrangean level. Moreover, further complicated non-local
constraints associated with the choice of a gauge-fixing condition (e.g.,
light-cone-gauge) also arise. These constraints, obtained by defining the
currents
\bearst
J_- (W)&=& {1\over 4\pi} Wi\partial_- W^{-1}\quad ,\quad
J_+ (W) = {1\over 4\pi} W^{-1}i\partial_+W\\
j_-&=&\psi_1^{(0)}\psi_1^{(0)\dagger}+\lbrace b_-^{(0)},c_-^{(0)}\rbrace
\quad , \quad
j_+=\psi_2^{(0)}\psi_2^{(0)\dagger}+\lbrace b_+^{(0)},c_+^{(0)}\rbrace\quad ,
\eearst
are as follows
\bearst
\Omega_+&=&-(1+c_V)J_+(\tilde W)+j_+, \\
\Omega_-&=&-(1+c_V)J_-(\tilde W')+j_-\quad ,\\
{\hat\Omega}_-&=&-\lambda^2\beta(\partial_+^{-2}(\beta^{-1}
i\partial_+\beta))\beta^{-1}+J_-(\beta)\\
&&-(1+c_V)J_-(\tilde W)+\lbrace {\hat b}_-^{(0)},\hat c_-^{(0)}
\rbrace\quad .
\eearst

The physical Hilbert space can be obtained in terms of these constraints.
Notice that the fields $\beta$ are not 
physical themselves. The $\beta$ action itself describes an integrable
model, but due to quantum corrections the $\beta$ excitations are in principle
not physical, while the physical
mesons do not describe an integrable model, albeit they are almost stable
against decay.\cite{abd-moha}

The massive theory, although not factorizable (even only at the lagrangean 
level) can be rewritten in terms of fermions, and the question of 
screening can be discussed.

We understand the phenomenon with a pair of probe charges $(q,-q)$, 
a distance $L$ apart, leading to the current
\bearst
J^0=q\left[ \delta(x-L/2)-\delta(x+L/2)\right] =-{e\over \sqrt\pi}
{\partial Q\over \partial x}
\eearst

Suppose the colour is fixed, e.g. $q\sim \sigma_2\in SU_2$.

The effective Lagrangian of the equivalent mechanical problem
is
\bearst
{\cal L}&=&{1\over 2}{E'}^2 +{1\over 2}{c_V+1\over c_V}{\Phi '}^2
+2\pi\lambda^2E^2 -2m^2\cos 2\sqrt\pi (\Phi +E) +\\
&& 2\sqrt\pi (c_V+1)\lambda q E [\theta (x+L/2) -\theta (x-L/2) ]\\
&&-{1\over 2} (c_V +1 )^2 q^2  [\theta (x+L/2) -\theta (x-L/2) ] 
-{{\psi '}^2\over 2c_V}
\eearst
where
\bearst
\Phi&=&\varphi +\eta\quad ,\quad \psi =\varphi +(c_V+1)\eta\quad ,\\
g&=& \E^{2i\sqrt\pi\varphi\sigma_2}\quad ,\quad
\beta = \E^{2i\sqrt\pi E\sigma_2}\quad ,\quad
\Sigma =\E^{2i\sqrt\pi\eta\sigma_2}\quad .
\eearst

The combinations
\be
\chi_+\approx E-a\Phi\quad \hbox{and  } \quad \chi_-\approx \Phi
-\epsilon aE
\ee
are described by the lagrangian
\be
{\cal L}=-{1\over 2c_V}{\psi '}^2 +\sum_{\pm}
\left[ {1\over 2}{\chi '}^2_i +{1\over 2}m_i\chi^2_i +
\lambda Q_i\chi_i\right]\quad .
\ee
We see above that there are two parameters describing mass, and one expects
already at this point a screening like potential.

Solving the classical problem (see Schwinger model) we obtain
\be
V(L)= A{1-\E^{-m_+L}\over m_+} + B {1-\E^{-m_-L}\over m_-}
\ee
which always leads to screening.

Notice that $m_-\sim c_V$, and for the abelian case one always gets a
confinement terms, unless $B=0$.

\section{Applications to four dimensional QCD}

The issue of quasi-integrability, by which we mean a rather weakened version
appears in four dimensional QCD. The case of high energy scattering and
its relation to the Heisenberg model is already well know and
sufficiently discussed.

Let us first review methods to deal with perturbative computations in
QCD, a task whose difficulty is enhanced by the number of diagrams,
colour indices, and polynomial vertices.

There are evidences for a weak version of integrability of the Yang-Mills 
theory. These resemble those of
two-dimensional Yang-Mills theory, where integrability seems to fail
by small deviations, when probed by means of the meson decay amplitudes.

Available methods for treating four dimensional QCD are:

$\bullet$ Helicity amplitudes \hfill\break
one deals with external particles with definite helicity. In the high energy,
or massless fermion, limit one has maximal helicity violation.

$\bullet$ Colour decomposition \hfill\break
use of group theory to break amplitudes into gauge invariant pieces,
with a fixed cyclic order of external legs. Useful for obtaining amplitudes
which contain a large number of external gluons.

$\bullet$ Supersymmetry identities\hfill\break
although QCD is not supersymmetric, at tree level
there being no fermion loops, the is no loss considering
the fermions in the adjoint representation. Thus, supersymmetry
can be used to relate the amplitudes containing gluons to others
containing scalars, which leads to significant simplifications. Usefull 
at one loop level.

$\bullet$ Recurrence relations  \hfill\break
Amplitudes with one off-shell quark or gluon can be obtained using 
classical solutions of Yang-Mills equations.  More specifically,
knowing an amplitude with $n$ external on-shell legs, one can use 
these relations to compute amplitudes with $(n+1)$ on-shell legs.

$\bullet$ String inspired methods  \hfill\break
There are consistent string theories whose infinite-tension limit 
corresponds to a non-abelian gauge field. In such cases, one can do 
loop computations, by using the string formulation, as a bookkeeper of
the algebra.

$\bullet$ Unitarity \hfill\break
use of Cutkovsky rules and analytic methods are also 
useful in simplifying the results.
\subsubsection{Helicity amplitudes}

Some results are extraordinarily simple, in spite of the difficulty of the 
subject.

Thus tree-level helicity amplitudes are such that if all helicities, or all 
but one, are the same, the amplitude vanishes. If only two are different, the
tree amplitude will be simply:
\bearst
{\cal A}(p_1,+,p_2,+,\cdots p_n,+)&=&0\\
{\cal A}(p_1,-,p_2,+\cdots p_n,+)&=&0\\
{\cal A}(p_1,-,p_2,-,\cdots p_n,+)&=&i e^{n-2}{\langle 12\rangle ^4\over
\langle 12\rangle\langle 23\rangle\cdots\langle (n-1)n\rangle
\langle n1\rangle}
\eearst

One loop results look like anomalies,
\bearst
{\cal A}(p_1,+,p_2,+,p_3,+,p_4,+)&=&i {e^4\over 2\pi^2}
{\langle 12\rangle^*
\langle 34\rangle^*\over \langle 12\rangle\langle 34\rangle}\\
{\cal A}(p_1,-,p_2,+,p_3,+,p_4,+)&=&i {e^4\over 2\pi^2}
{\langle 12\rangle
\langle 34\rangle^*\langle 24\rangle ^*\over \langle 12\rangle^*
\langle 34\rangle \langle 24 \rangle}
\eearst
\subsubsection{ Self-dual Yang-Mills theory}

Self-dual Yang-Mills theory implies very simple results for helicity 
amplitudes. In fact, tree-level amplitudes with a fixed helicity can
be given in terms of solutions of the self-dual Yang-Mills
theory. It is described by
\bearst
\left[ \gamma^\mu,\gamma^\nu\right] F_{\mu\nu} =0\; ,
\eearst
or
\bearst
F_{0+3,0-3}=F_{1+i2,1-i2}\quad F_{0-3,1-i2}=0=F_{0+3,1+i2}
\eearst
In the light-cone-gauge $A_-A_3=0$ we have a simple set of equations:
\bearst
G_{0+3,0-3}&=&-\partial_{0-3}A_{0+3}\\
G_{1+i2,1-i2}&=&\partial_{1+i2}A_{1-i2}-\partial_{1+i2}A_{1-i2}+i
\left[A_{1+i2},A_{1-i2}\right]\\
G_{0+3,1-i2}&=&\partial_{0-3}A_{1-i2}\\
G_{0+3,1+i2}&=&\partial_{0+3}A_{1+i2}-\partial_{1+i2}A_{0+3}+i
\left[A_{0+3},A_{1+i2}\right]
\eearst
whose solution is given in terms of a single scalar field, namely
\bearst
A_{1+i2}=\partial_{0-3}\Phi\quad , \qquad A_{0+3}=\partial_{1-i2}\Phi
\eearst
where $\Phi$ obeys the equation
\bearst
\partial^2\Phi + i \left[ \partial_{1-i2}\Phi,\partial_{0-3}\Phi\right]
\eearst

Such an equation of motion may be derived from the lagrangean
\bearst
{\cal L}& =& \varphi\left[ 
\dal\Phi +i\partial_{1-i2}\Phi,\partial_{0-3}\Phi
\right]\\
& =& \varphi\left[ 
\dal\Phi +i\partial_+^{\dot\alpha}\Phi,\partial_{+\dot\alpha}\Phi
\right]
\eearst
The field $\varphi$ has dimension 2 and cannot make an external leg
except at tree level. Moreover, due to the dimension and
helicity counting, all amplitudes are maximally helicity violating.
Indeed, we find the previous amplitudes. 

An iterative solution can be found in terms of the 
colour-ordered amplitude expansion. We have

\bearst
&&\Phi\sim \phi (k_1)e^{-ik_1x} T^{a_1} +\\
&& +i \sum 
\phi (k_1)e^{-ik_1x} T^{a_1}\cdots \phi (k_n)e^{-ik_nx} T^{a_n}
(Q_1-Q_2)^{-1}(Q_2-Q_3)^{-1}\cdots (Q_{n-1}-Q_n)^{-1}
\eearst
with $Q=k_{0+3}/k_{1-i2}$.

As noticed by Bardeen, the form of the solution corresponds to the (ordered)
Bethe-Ansatz solutions found in two-dimensional integrable systems.

Note however: Coleman-Mandula theorem is a no-go for integrability in
higher dimensional systems!!!
\section{A note on conformal invariance}
The use of conformal symmetry in quantum field theory has been advocated
by Wess and Kastrup at the beginning of the sixties.
Renewed interest in this topic resulted from Wilson's ideas about the
short distance expansion of products of operator at nearby points, and 
the associated
notion of anomalous dimensions of field operators, which are
intimately linked to the high energy behaviour of renormalizable
quantum field theories.

It was noted by Polyakov and others in the early seventies,
that critical models implement a global conformal invariance which
goes beyond pure scale invariance. This has led Polyakov to  propose
the use of conformal invariance as an essential ingredient in the
study of the critical behaviour of statistical models at second order phase
transitions.

Whereas scale transformations merely scale relative distances by a
constant factor, conformal transformations involve a space-dependent
factor. This imposes restrictions, which allow one to
fix the two- and three-point functions at criticality. In general one
is not able to go beyond that, since in general the conformal group is
a finite dimensional Lie group,  resulting in only a finite number
of restrictions on the correlators.

In two dimensions the situation is drastically different, since here the
conformal transformations are represented by all the analytic 
transformations in euclidean space. This will in fact enable us to reduce 
the two-dimensional problem to two one-dimensional ones. In fact, the
significant restrictions imposed by the invariance under analytic
transformations will ultimately lead to a classification of a large class of
critical phenomena in two dimensions.

Conformal invariance also finds its application in string
theory, two-dimensional gravity, as well as the non-linear sigma
models.

\subsection{The conformal group in two dimensions}
In two dimensions the Killing-Cartan equation takes the form
\be  
\partial_\mu\epsilon_\nu+\partial_\nu\epsilon_\mu=(\partial\cdot\epsilon)
\eta_{\mu\nu}\label{killingcartan2}
\ee 
The diagonal elements of these equations then read
$\partial_0\epsilon_0+\partial_1\epsilon_1=0$,
whereas for the off-diagonal elements we have
$\partial_0\epsilon_1+\partial_1\epsilon_0=0$.
From these equations it follows that
\be  
(\partial_0\pm\partial_1)(\epsilon_0\pm\epsilon_1)=0
\label{plusminusdereps}
\ee 
Hence in $D=2$, a general conformal transformation is parametrized
by $\epsilon_\pm =\epsilon_0\pm\epsilon_1$, where $\epsilon_+(\epsilon_-)$
depends only on the light-cone-variable $x^+=x^0+x^1(x^-=x^0-x^1)$.
The implications of this fact can best be appreciated by going
to euclidean space via the usual substitution $x^0=-ix_2^E$, $x^1=x_1^E$,
with the corresponding substitution $\epsilon ^0=-i\epsilon_2^E$,
$\epsilon^1=\epsilon_1^E$. The Cartan-Killing  equations now read
\bear
\partial_\mu\epsilon_\nu^E+\partial_\nu\epsilon_\mu^E=
(\partial^E\cdot\epsilon^E)\delta_{\mu\nu}\nonumber
\eear
implying\\
\ptwelve{\bearst
\partial_1^E\epsilon_1^E-\partial_2^E\epsilon_2^E&=&0\\
\partial_2^E\epsilon_1^E+\partial_1^E\epsilon_2^E&=&0\eearst}
\hfill\peight{\bear\label{cauchyriemann}\eear}\\
If we define the complex function
\bear
\epsilon=\epsilon_1^E+i\epsilon_2^E\nonumber
\eear
and the complex variable
\bear
z=x_1^E+ix_2^E\nonumber
\eear
then equations  (\ref{cauchyriemann})  are just the Cauchy-Riemann equations
for the real  and imaginary  parts of the function $\epsilon$,
\be 
{{\partial Re \epsilon}\over{\partial x_1^E}}={{\partial Im\epsilon}
\over{\partial x_2^E}},\quad
{{\partial Im\epsilon}\over{\partial x_1^E}}=-{{\partial Re\epsilon}
\over{\partial x_2^E}}\quad .\label{cauchyriemann2}
\ee  
This implies that the function $\epsilon$ depends only on the complex
variable  $z$, and $\epsilon^*$ on the complex variable  $z^*=\overline z$:
\be
\epsilon_1^E+i\epsilon_2^E=\epsilon(z);\quad\epsilon_1^E-i\epsilon_2^E
=\overline\epsilon(\overline z)\label{holo}
\ee 
Evidently, we have $\overline\epsilon (z^*)=\epsilon^*(\overline z)$.
The two dimensional conformal transformations thus coincide
with the analytic transformations
\be  z\to f(z),\quad \bar z\to\bar f(\bar z)\quad .\label{holotransf}\ee 
We have the following useful relations
\bear  \partial_z&=&{1\over2}(\partial_1^E-i\partial_2^E)\; ,\quad 
\partial_{\bar z}={1\over2} (\partial_1^E+i\partial_2^E)\nonumber\\ 
b\cdot x^E&=&{1\over2}(b\bar z+\bar bz)\quad
b\cdot\partial^E=(b\partial_z+\bar b\partial_{\bar z})\nonumber
\eear
where $b=b_1+ ib_2$  $\bar b = b_1-ib_2$.

In the new coordinates the element of length and
metric are respectively given by 
\be
ds^2=dzd\bar z\; ,\ \eta_{z\bar z}=\eta_{\bar zz}={1\over2}\; ,\ \eta_{zz}
=\eta_{\bar z\bar z}=0 \quad .\label{newcoord}
\ee 
Therefore, under a change of variable $z\to z'$ defined by  $z = f(z')$,
\be
ds^2 \to\left({{\partial f}\over{\partial z}}\right)
\left({{\partial\bar f}\over
{\partial \bar z}}\right)dzd\bar z\; .\label{transmetric}
\ee

Since $f(z)$ and $\bar f(z)$ are respectively analytic functions of $z$ 
and antianalytic of $\bar z$, 
they have a Laurent expansion of the form
\be 
f(z)= z+\sum^\infty_{-\infty}\epsilon_n z^{n+1}\; ,\quad
\bar f(\bar z)=\bar z+\sum^\infty_{-\infty}\bar \epsilon_n \bar z^{n+1}\; .
\label{fzzbar}
\ee 

We may regard this expansion as an expansion in the basis functions
\bear
\psi_n=z^{n+1},\quad \bar\psi_n=\bar z^{n+1}\nonumber
\eear
and write
\bear
z'=\left(1+\sum^\infty_{-\infty}\epsilon_nL^c_n\right)z\quad
\overline z'=\left(1+\sum^\infty_{-\infty}\epsilon_nL^c_n\right)\bar z\nonumber
\eear
where  $L_n^c=z^{n+1}{d\over{dz}}$,
${\bar L}_n^c={\bar z}^{n+1}{d\over{d\bar z}}$ 

The operators $L^c_n$ and  ${\bar L}^c_n$ are the generators of the analytic
transformations and satisfy the loop algebra
\bear 
\lbrack L^c_n,L^c_m\rbrack &=&-(n-m)L^c_{n+m}\; ,\nonumber\\
\lbrack\bar L^c_n,\bar L^c_m\rbrack &=&-(n-m)\bar L^c_{n+m}\; ,
\label{loopalgebra}\\
\lbrack L_n^c,\bar L_m^c\rbrack &=&0\; .\nonumber
\eear 
The superscript $c$ is used to remind the reader that the $L_n^c$ generate
analytic transformations on classical functions.
In the quantum case the first two commutators will be modified
by an extension proportional to a central charge, while the third commutator 
remains unchanged. This defines the Virasoro  algebra.

Since the  $L_n^c$   commute with the  $\bar L_m^c$   , the local  conformal 
algebra  is the direct sum $A \oplus \bar A$
of the two isomorphic algebras.
This implies that  the conformal group in two dimensions
acts independently on $z$  and  $\bar z$.
For this reason we
may continue the Green functions of a conformal
theory in $D = 2$ to a
larger domain, where $z$ and $\overline z$  are treated as independent
variables, as already advertised. By taking $\overline z$ as the complex 
conjugate of $z$, we recover the original coordinates $(x_1,x_2)\in \Real^2$.

\chapter{Two-dimensional Quantum Chromodynamics}
\section{Introduction}
The Schwinger model, or QED in two-dimensions with massless
fermions, is exactly solvable. The physical space is generated by a single
massive free boson, while the original fermions are bounded together by a
screening force, such that the charge quantum number is not observable. In
the massive fermion case,
a long range force develops, and further quantum numbers associated with the
fermion are permanently confined. The vacuum is degenerate, an issue which can
be traced back to the break of chiral symmetry. A plethora of interesting
physical results, expected to be true in a realistic parton model but which
had never been firmly established, were proved\cite{aar}.

Unlike the Schwinger model, quantum chromodynamics of massless
fermions is no longer exactly solvable.
It nevertheless serves as a very useful laboratory for studying
problems such as the bound-state spectrum, algebraic structure and duality
properties. These problems important tools for general understanding of
realistic quantum field theories and are expected to be realized in four
dimensional Quantum Chromodynamics. In particular, exact properties may 
be derived, by either using standard perturbation theory, or by obtaining 
its effective action using the heat kernel method. One arrives at an 
equivalent bosonic formulation in the form of a gauged Wess-Zumino-Witten 
(WZW) action.

The first attempt to obtain the particle spectrum dates back to 1974, and was
based on the $1/N$ expansion\cite{thooft}, where $N$ is the number of colours.
In this limit one is led to a bound state spectrum corresponding asymptotically
to a linearly rising Regge trajectory. The use of the principal-value 
prescription in dealing with the infrared divergencies is however 
highly ambiguous due to the non-commutative nature of principal-value 
integrals. Moreover, the result for the fermion propagator is 
tachyonic for a small fermion mass as compared with the coupling constant,
hence, in particular, for $m=0$. This has made 't Hooft's solution a 
controversial issue\cite{review}.

In the large $N$ approximation the gluons remain massless, since fermion 
loops do not contribute to the Feynman amplitudes. This is unlike the 
$U(1)$ case, where the 
photon acquires a mass via an intrinsic Higgs mechanism. This has led to 
the speculations that $QCD_2$ may in fact exist in two phases associated with
the weak and strong coupling regimes. In this picture, the large $N$ limit
would correspond to the weak-coupling limit ('t Hooft's phase), with
massless gluons and a mesonic spectrum described by a Regge trajectory. 
In such a case, the Regge behaviour of the mesonic spectrum is compatible
with confinement. In the strong coupling regime (Higgs
phase), on the other hand, the gluons would be massive, and the original
$SU(N)_c$-symmetry  would be broken down to the maximal abelian subgroup
(torus) of $SU(N)_c$.

The behaviour of the theory in the strong or weak coupling limits is
rather subtle. The theory
is asymptotically free, as it should since it is super-renormalizable. In the 
strong coupling limit, it is expected to be in the confining phase.
Indeed, in the infinite infrared cut-off limit  quarks disappear from 
the spectrum, which consists of mesons lying approximately on a 
Regge trajectory. One obtains a simple and finite solution  for the fermion
self-energy (SE), and the fermion two-point function\cite{cacoogr}. 
This solution is useful for analyzing properties related to the high-energy 
scattering amplitudes.

Functional techniques are also very powerful for
arriving at equivalent bosonic actions for the fermionic theory, and
for obtaining a non-abelian bosonization formulae.

In the framework of conventional perturbation
theory, by formally summing the perturbation series,  we arrive at
exact representations for the gauge current, the determinant of the Dirac
operator  and the Dirac Green's functions  in an external field. 
The external field current $J_\mu (x\mid A)$ and fermionic determinant 
$\det i\not \! \! D$ of a $U(1)$ gauge theory in two 
dimensions $(QED_{2})$ are exactly calculable. In fact they  are, respectively,
functionals of first and second degree in $A_\mu $, reflecting the
fact, that the perturbation series for $J_\mu (x\mid A)$ and $\ln \det \; i
\not \! \! \! D[A]$ (connected one-loop graphs) terminates at the first 
non-trivial order in the coupling constant $e$. Subsequently, we could 
also solve for the exact Dirac-Greens functions in this case. Moreover, the 
structure of both the determinant and the Greens function is such that all 
correlation functions of $QED_{2}$ can be constructed explicitly via the 
generating functional for the massless case.

The same is not true for $QCD_{2}$, where the perturbation series for
$\ln\det i\!\not\!\!\! D$, corresponding to the effective action and $J_\mu
^a(x\mid A)$ no longer terminates. Nevertheless, the effective action 
may still be calculated non-perturbatively in closed form by the Fujikawa 
method, or the heat-kernel (proper-time) method, by integrating the anomaly
equation, or finally by summing the perturbative series. The result
can be expressed in terms of a gauge invariant combination of the gauge 
potentials in the form of a Wess-Zumino-Witten action.

We shall primarily be dealing with massless fermions. The massive
fermion case can in general not be dealt with, if one seeks exact results. 
It is however possible to compute the functional determinant as an expansion 
in the inverse of the mass. The problem of screening and confinement can
however be analysed along the lines of the $U(1)$ case. However, unlike the
$U(1)$ case, the screening phase prevails in the non-abelian 
theory\cite{gross-kle-etc,amz}.
\section{The Wess-Zumino-Witten theory}
That a bosonic action can be used to describe baryons was foreseen
long time ago\cite{44}: it has been known that the baryon 
number of the solution of a certain bosonic theory is non-zero.
Moreover, Lagrangeans of the type of a sigma 
model containing a topological term provided a rather reasonable 
description of low energy hadron physics.

In two-dimensional space-time, we aim at an equivalent bosonic
action for massless fermions, transforming under a non abelian
symmetry group. One is thus led to consider the two dimensional action
\be
\Gamma[g]\equiv S_{WZW}=S_{P\sigma M}+S'\quad ,\label{s-wzw=spsig+sprime}
\label{wznwaction}
\ee
where
\be
S'=n S_{WZ}={n\over 4\pi }\int _0^1 dr \int d^2x \epsilon ^{\mu \nu 
}tr  g_r^{-1}\partial _rg_rg_r^{-1}\partial _\mu g_rg_r^{-1}\partial _\nu
g_r\label{wess-zumino-term}
\ee
and $g_r(r,x)$ extends $g(x)$ in such a way that $g_r(0,x)=1$ and 
$g_r(1,x)=g(x)$. The principal sigma model action reads
\be
S_{P\sigma M}={1\over 2\lambda^2 }\int d^2xtr\, 
\partial ^\mu g^{-1}\partial _\mu g\quad .\label{sprisigma}
\ee
We shall refer to $S_{WZ}$, $S_{P\sigma M}$ and $S_{WZW}$ as the 
Wess-Zumino (WZ), principal sigma  model ($P\sigma M$),
and Wess\-Zumino\-Witten  ($WZW$) action,  respectively\cite{wittencmp}.
\subsubsection{Existence of a Critical Point}
\noindent Consider the one parameter family of actions (\ref{wznwaction}),
to which shall refer as the Wess\-Zumino\-Novikov\-Witten
(WZNW) action. For the choice ${4\pi 
\over \lambda ^2}=n$, $S_{WZNW}$ is the equivalent bosonic action 
of the (conformally invariant) theory of $N$ non-interacting massless 
 fermions. To demonstrate this we examine the equations of
motions and one-loop $\beta $-function corresponding to (\ref{wznwaction}).

The equations of motion are obtained by computing the functional 
variation of $S_{WZNW}$ with respect to $g_{ij}$. From 
(\ref{s-wzw=spsig+sprime}-\ref{sprisigma}) we obtain two alternative 
forms for $\delta S_{WZNW}$ (see \cite{ar-prd} for details):
\bear
\delta S_{WZNW}&=&\int d^2x \tr g^{-1}\delta g\left( {1\over
\lambda  
^2}g^{\mu \nu }-{n\over 4\pi}\epsilon ^{\mu \nu }\right) \partial _\mu 
(g^{-1}\partial _\nu g)\quad ,\\
&=&\int d^2x tr \delta g g^{-1}\left( -{1\over \lambda ^2}g^{\mu \nu 
}-{n\over 4\pi}\epsilon ^{\mu \nu }\right) \partial _\mu  \left( g\partial _\nu
g^{-1}\right)\quad .
\eear
where the integration over $r$ could be performed.

Setting $\delta S_{WZNW}=0$, we obtain as equations of motion the 
conservation laws,
\bear
\partial _\mu &&\left( {1\over \lambda ^2}g^{\mu \nu }-{n\over 4\pi }
\epsilon ^{\mu \nu }\right) (g^{-1}\partial _\nu g)=0\label{cons1-wznw}\\
\partial_\mu&&\left( {1\over \lambda ^2}g^{\mu \nu }+{n\over 4\pi }\epsilon 
^{\mu \nu }\right) (g\partial _\nu g^{-1})=0\quad .\label{cons2-wznw}
\eear

For the choice
\be
{4\pi \over \lambda ^2}=n\quad ,
\ee
Eqs. (\ref{cons1-wznw}-\ref{cons2-wznw}) just express the conservation of
the left ($L$) and right ($R$) moving currents 
\bear
j_L^\mu (x)& =&-i{n\over 4\pi }(g^{\mu \nu }-\epsilon
^{\mu \nu  })(g^{-1}\partial _\nu g)\quad ,\quad
j_R^\mu (x)=-i{n\over 4\pi }(g^{\mu \nu }+\epsilon ^{\mu \nu 
})(g\partial _\nu g^{-1})\quad .\label{critcurr1}\\
j_+(x)&=&-{in\over 2\pi }g^{-1}\partial _+g=i2\hat \Pi
^Tg-{in\over 2\pi }g^{-1}\partial _1g\quad ,\label{critcurr2}\\
j_-(x)&=&-{in\over 2\pi }g\partial _-g^{-1}=-i2g\hat \Pi
^T+{in\over 2\pi }g\partial _1g^{-1}\quad ,\label{critcurr3}
\eear
This indicates that at the critical point  the WZNW 
action is a natural candidate for describing $N$ non-interacting 
massless fermions.

\subsection{Properties at the Critical Point: the Kac Moody algebra}

From the above discussion, we expect the theory to have a conformally
invariant fixed point. To investigate the properties of the WZNW action
(at the critical point) further,
we proceed with a canonical quantization of the theory.

The WZ action  depends linearly on the time derivative of 
$g$. Hence it proves useful to write it in the form\cite{ar-prd}
\be
S'[g]={n\over 4\pi } \int d^2x\, tr A(g)\partial _0g\quad ,
\ee
where we have formally integrated over $r$, and $A(g)$ is a 
matrix-valued function of $g$ and $\partial _1g$. We thus obtain for the 
momentum conjugate to $g_{ij}$
\be
\Pi _{ij}={n\over 4\pi }\partial _0g^{-1}_{ji}+{n\over 4\pi
}A_{ji}\equiv \hat \Pi _{ij}+{n\over 4\pi }A_{ji}\quad .
\ee
Hence there are no constraints, and we can compute the Poisson Brackets,
obtaining
\bear
\left\{ j_+^a(x),j_+^b(y)\right\}&=&2 f^{abc}j_+^c(x)\delta
(x^+-y^+)+ {2n\over \pi }\delta ^{ab}\delta '(x^+-y^+)\label{pbcurrents0}\\
\left\{ j_-^a(x),j_-^b(y)\right\}&=&2 f^{abc}j_-^c(x)\delta (x^-y^-)+
{2n\over \pi }\delta ^{ab}\delta '(x^-y^-)\label{pbcurrents}\\
\left\{ j_+^a(x),j_-^b(y)\right\}&=& 0\quad .\label{pbcurrents2}
\eear
with $j^a=\tr j\tau^a$, and $\tau^a$ the $SU(N)$ generators. 
Here ``prime" on $\delta $ represents the derivative with respect
to the argument of $\delta$. Thus, we have two  Kac-Moody algebras with central
extension $k=n$.

It is easy to show that (\ref{pbcurrents0}-\ref{pbcurrents2}) defines
also the algebra one 
obtains in a field theory of $N$ free  fermions, with the action 
$S={i\over 2} \int d^2x\overline \psi ^i\not \! \partial \psi ^i$.

Since a Kac-Moody algebra defines a conformally
invariant theory uniquely for $k=\pm 1$ (unitary irreducible representations
are unique), we are led to identify the currents 
(\ref{critcurr1}-\ref{critcurr3}) with $ \tr 
(\tau^aj_\pm )$. To complete this picture, we need to identify both 
the fermionic representation of the bosonic field $g_{ij}$, 
and the energy-momentum tensor in these theories. 

It is clear that the WZW action corresponds to an interacting conformally
invariant field theory for general values of $k$. It can  be used in the
bosonisation procedure for higher representations of interacting fermions,
as in the case of QCD$_2$.

We can now prove the following important result due to 
Polyakov-Wiegman\cite{50}.

{\it Theorem}
\be
\Gamma[AB]=\Gamma[A]+\Gamma[B]+{1\over 4\pi }\int
d^2x(g _{\mu \nu }+\epsilon _{\mu \nu }){\rm tr} (A^{-1}\partial
_\mu A)(B\partial _\nu B^{-1}).
\ee

{\it Proof}: It is convenient to use the notation
\be
\Gamma[G]=S_{P\sigma M}[G]+S_{WZ}[G]\quad ,
\ee
where $S_{P\sigma M}$ is the principal sigma  
model action, and the  functional $S_{WZ}[G]$ is the Wess-Zumino action 

Replacing $G_r(G)$ by $A_rB_r(AB)$  and using the cyclic properties
of the trace one readily finds  
\be
S_{P\sigma M}[AB]=S_{P\sigma M}[A]+S_{P\sigma M}[B]+{1\over
4\pi }\int d^2x {\rm tr} [(A^{-1}\partial _\nu A)(B\partial _\mu
B^{-1})].
\ee

The corresponding calculation
for the Wess-Zumino term is more involved.
Using $(AB)^{-1}\partial (AB)=B^{-1}[A^{-1}\partial A+(\partial B)B^{-1}]B$, 
one finds after a lengthy calculation, that 
\be
S_{WZ}[AB]=S_{WZ}[A]+S_{WZ}[B]-{1\over
4\pi }\int _0^1dr \int d^2x \epsilon _{\mu \nu }{\cal W}_{\mu \nu }\quad ,
\ee
where ${\cal W}_{\mu \nu }$ can be
written in the form 
\be
{\cal W}_{\mu \nu }={d\over dr} {\rm tr} [(A^{-1}\partial _\mu
A)(B\partial _\nu B^{-1})]
-\partial _\mu [(B\partial _\nu B^{-1})A^{-1}\dot A]-\partial
_\nu [(A^{-1}\partial _\mu A)B\dot B^{-1}]\quad .\label{wmunu}
\ee

The second and third term in (\ref{wmunu}) contribute a surface
term to $S_{WZ}[AB]$, which we drop. The first term gives a
contribution which can be trivially integrated in $r$. Hence
\be
\Gamma [AB]=\Gamma [A]+\Gamma [B]+{1\over
4\pi }\int d^2x {\rm tr} [(A^{-1}\partial _+ A)(B\partial _-
B^{-1})]\quad ,\label{pol-wieg-id}
\ee
which is the Polyakov-Wiegmann identity.
\section{QCD$_2$:currents and functional determinant.}
Quantum chromodynamics is defined by the Lagrange density
\be 
{\cal L} = -{1\over 4}\tr F_{\mu\nu}F^{\mu\nu} + \overline \psi_i (i \!\not \! 
\partial + e \!\not \!\! A)\psi_i \quad ,\label{qcd2lagrangian}
\ee 
where $F^{\mu\nu}$ is the chromoelectric field tensor, and the fermions 
$\psi_i$ are in the fundamental representation of the gauge group.

The current $J_\mu^a = \overline \psi \gamma_\mu \tau ^a \psi$ is covariantly 
conserved as a consequence of the Dirac equation. For a gauge-invariant 
regularization such a conservation law holds after quantization.
We consider in general an external-field current, which is 
obtained by differentiating the functional
\be 
W[A]=-i\, \ln{\det i\not\!\!D[A]\over\det i\not\!\partial}\quad ,
\label{funcdeterminant}
\ee 
with respect to $A_a^\mu$, and is given by the expression
\be 
e J_\mu^a(x\vert A) =  {\delta W\over \delta A_a^\mu(x)}\quad .
\label{extfieldcurrent2}
\ee 
The functional $W[A]$ represents an effective action for the gauge field
$A_\mu$.

In 1+1 dimensions the fermionic part of the Lagrangian (\ref{qcd2lagrangian})
is classically invariant under both $U(1)$ and chiral gauge transformations.
For a $U(1)$ gauge invariant regularization, the local chiral symmetry 
is broken at the quantum level. Following Fujikawa, we can view this fact
as the non-invariance of the fermionic measure under a local 
chiral change of variables. The corresponding Jacobian is obtained from the 
anomalous behaviour of the effective action under this transformation. With 
the definition of the axial vector current depending on the external gauge
field as $J_{5\mu}^{b}(x\vert A)=\epsilon_{\mu\nu} J^{\nu b}(x\vert A)$,
one thereby obtains from (\ref{extfieldcurrent2}) the  anomaly equation
\be
{\cal D}_\mu^{ab}{J^{5\mu}}^b=-\widetilde{\cal D}_\mu^{ab} 
{J^\mu}^b={e\over 2\pi}
\epsilon_{\mu\nu}{F^{\mu\nu}}^a\quad .\label{anomalyequation}
\ee 
where ${\cal D}$ is the gauge covariant derivative in the adjoint
representation.

The above anomaly equation leads to the
pseudo-divergence of the Maxwell equation, that is, 
\be 
\epsilon_{\nu\rho}{{\cal D}}^\rho {{\cal D}}_\mu F^{\mu\nu} + e \widetilde 
{{\cal D}}_\mu J^\mu = - {1\over 2}\left({{\cal D}}^2 + {e^2\over \pi} \right)
\epsilon_{\mu\nu}F^{\mu\nu}=0\quad ,\label{divanomalyeq}
\ee 
showing a mass generation for the gauge field, analogous to the one in 
the Schwinger model.

Furthermore, it is possible to compute the external field current
$J_\mu^a (x\vert A)$ by integrating (\ref{anomalyequation}). Using the kernel 
$K_\mu^{ab} (x,y\vert A)$ defined by the equations
\be 
{{\cal D}}_\mp^{ab} K_\pm^{bc} = \mp\delta^{ac}\delta (x-y)\quad , 
\label{kernelequation}
\ee 
we have 
\be
J_\pm ^a = {e\over 2\pi} \int {\rm d}^2y\, K_\pm^{ab}(x,y\vert A) 
\epsilon_{\rho \sigma}{F^{\rho\sigma}}^b(y)  \quad .\label{rescurrentpm}
\ee
Defining the two-point functions $D_\pm=-{i\over 4\pi x^\mp}$
we may represent $K_\mu$ as a power series expansion in terms of the
gauge field in the fundamental
representation,  $A_\mu = \sum _c A_\mu ^c \tau^c$. One finds
\bear
K_\pm ^{ab}(x,y\vert A) &=& \,  \delta^{ab}D_\pm (x-y) - \nonumber\\ 
&-& \sum _{n=1}^\infty (-e)^n\int {\rm d}^2x_1 \cdots {\rm d}^2x_n \, D_\pm 
(x-x_1)\cdots D_\pm (x-x_n)\nonumber\\ 
& \times& \tr \left\{ \tau^a \left[ A_\mp (x_1), [A_\mp (x_2), \cdots [A_\mp 
(x_n),\tau^b ]]\cdots \right]\right\} \quad .\label{kernelresultsfund}   
\eear 

Substituting (\ref{kernelresultsfund}) into
Eq. (\ref{rescurrentpm}) and making use of the fact that
the gauge field strength $F_{+-}$ may be alternatively written in the form
$F_{+-}=-\partial_-A_+ 
+{\cal D}_+A_-$ or $F_{+-}=-{\cal D}_-A_++\partial_+A_-$, one obtains, after 
a partial integration (and use of (\ref{kernelequation}),
\bear
J_\pm^a(x\vert A) &=& {e\over 2\pi} A_\pm (x) - {e\over 2\pi}\int {\rm d}^2 y\,
K_\pm ^{ab}(x,y\vert A) \partial _\pm A^b_\mp  \nonumber\\ 
&=&{e\over 2\pi} \left[ A_\pm - \int {\rm d}^2y\, \partial _\pm D_\pm (x-y)
A_\mp^a(y)\right]\nonumber\\ 
&+& {i\over 2\pi} \sum _{n=2}^\infty (-ie)^n\int {\rm d}^2x_1 \cdots 
{\rm d}^2x_n \, D_\pm (x-x_1)\cdots D_\pm (x-x_n)\nonumber\\ 
&\times& \tr \left\{ \tau^a \left[ A_\mp (x_1), [ \cdots  [A_\mp (x_{n-1}),
\partial _\pm A_\mp (x_n)]] \cdots \right]\right\} 
\quad .\label{currentfunctionalpert}
\eear 
From (\ref{currentfunctionalpert}) we now compute the effective action
by integrating (\ref{extfieldcurrent2}) 
\bear
W[A]&=&W[0]-{ie^2\over 2\pi}\int {\rm d}^2 x\, \delta^{ab} A_\mu^{a}\left( 
g^{\mu\nu} - {\partial ^\mu \partial^\nu\over\partial^2}\right)A_\nu^b (x)
\nonumber\\ 
&+& {i\over 2}\sum_{n=2}^\infty {(ie)^{n+1}\over n+1} \int {\rm d}^2x\, \tr \,
\left[ A_-(x) T_+^{(n)} (x\vert A) + A_+(x) T^{(n)}_-(x\vert A)\right]\quad , 
\label{effectiveaction}
\eear
where
\bear
T_\pm ^{(n)} (x\vert A) &=& - {1\over 2\pi} (-1)^n \int {\rm d}^2x_1 \cdots 
{\rm d}^2x_n \, D_\pm (x-x_1)\cdots D_\pm (x-x_n)\nonumber\\ 
&\times&\left[A_\mp(x_1),[\cdots [A_\mp (x_{n-1}), \partial_\pm A_\mp (x_n) ]] 
\cdots \right] \quad . \label{tn}
\eear

The  exact summation of the series can be performed. We
first define the generating functional
\be 
\hat T_\pm(x,t\vert A)= \sum_2^\infty {(iet)^{n-1}\over (n-1)!}
T_\pm^{(n)}(x\vert A) \label{tofa}
\ee
such that 
\be 
T_\pm(x\vert A) =ie\int_0^1 dt T_\pm (x,t\vert A) \; . \label{tgener}
\ee 
Further we note that the functional $\hat T_\pm(x,t\vert A)$ 
satisfies the differential equation
\be \partial_\mp T_\pm = -{e\over 8\pi} \partial_\pm A_\mp + 
ie [A_\mp,T_\pm]\label{tdiffeq}
\ee
with $A_\pm (x,t)=tA_\pm (x)$. The above equation is easily solved by 
the expression
\be 
\hat T_\pm ={i\over 8\pi} \partial_\pm \hat U_\pm \hat U_\pm^{-1}(x,t)
\label{tresult}
\ee
where $\partial_\mp U_\pm =ie A_\mp U_\pm$. Replacing the gauge field 
by the $U$ fields defined above in the expression for $W$ 
(\ref{effectiveaction}), we find
\bear
W[A]&=& {1\over 4\pi}\int_0^1 dt\int d^2x \tr \partial_+U_+U_+^{-1}
\partial_t(\partial_-U_+U_+^{-1}) +\nonumber\\
&+& {1\over 4\pi}\int_0^1 dt\int d^2x \tr 
\partial_+U_-U_-^{-1}\partial_t(\partial_-U_-U_-^{-1}) +
{1\over 4\pi}\int d^2x \tr A_+A_-\quad .\label{perteffnonpert}
\eear

The first and second terms are of exactly the same form. The significance
of these terms can be understood by writing the first term as a sum of 
the symmetric 
plus anti-symmetric combinations in the $+$ and $-$ variables. The symmetric
part is readily seen to be a total derivative in the $t$ variable, and equals
the principal sigma model action for a matrix-valued field. The 
anti-symmetric part is handled by applying the $t$ derivative. When it acts 
on $U^{-1}$, the two terms obtained are exactly the topological term in 
the WZW action. When it acts on the derivative term, one can see that a
total derivative is obtained, that is $\partial_+[\partial_-U^{-1}
\partial_t U]-\partial_-[\partial_+U^{-1}\partial_t U]$. Therefore we 
obtain the sum of the WZW action for $U_+$, the corresponding one for 
$U_-$, and the contact term in $A_+A_-$, which according
to eq.(\ref{pol-wieg-id}) leads to the WZW action for the product
$U_+U_-$, which is the gauge invariant combination.
A similar result is obtained using other techniques\cite{aar}.
\subsubsection{The gauged WZW action }
As already remarked, the fermionic part of the Lagrangian 
(\ref{qcd2lagrangian}) is invariant under local gauge transformations 
$SU(N)$, as well as $SU(N)_L\times 
SU(N)_R$, for both right $(R)$ and left $(L)$ components, that is,
\bear
\psi _{\scriptstyle{R\atop L}} & \to &w_{\scriptstyle{R\atop L}}\psi
_{\scriptstyle{R\atop L}} \quad , \nonumber\\ 
A_\pm&\to &w_{\scriptstyle{R\atop L}}\left(A_\pm+{i\over e}\partial_\pm\right) 
w^{-1}_{\scriptstyle{R\atop L}}\quad .\label{rightleftgauge}
\eear
This transformation corresponds to pure vector gauge transformation when
$w_R=w_L=w$, while when
$w_R=w^{-1}_L=w$  it corresponds to a pure axial vector transformation. If we
use a change of variables parametrizing $A_\pm$ as
\be\label{aparametrization}
eA_+=U^{-1}i\partial_+U\quad ,\quad A_-=Vi\partial_- V^{-1}
\ee
the transformations corresponding to axial transformations reduce to $U\to 
U w^{-1}$ and $V\to w^{-1}V$. 

The above transformation is not a symmetry of the effective action $W[A]$ due
to the axial anomaly. This non-invariance may used in order to express the 
fermionic functional determinant in terms of a new
bosonic action $S_F[A,w]$ for  the fermions defined by
\be 
S_F[A,g]\equiv\Gamma[UgV^{-1}]-\Gamma[UV]\quad.\label{gaugedwzwg-g}
\ee
Using the invariance of the Haar measure, we evidently have, up to an
irrelevant constant,
\be 
\det i\not\!\!D\equiv\E^{iW[A]}=\int{\cal D}g\,\E^{iS_F[A,g]}\quad.
\label{detew}
\ee 

Thus $S_F(A,g)$ plays the role of an equivalent bosonic action for fermions
minimally coupled to gauge fields. Its
explicit form may be obtained by repeated use of the Polyakov-Wiegmann 
formula (\ref{pol-wieg-id}), and we obtain
\be 
S_F[A,g]=\Gamma[g]+{1\over 4\pi}\!\int\!{\rm d}^2x\,{\rm tr}\,\left[e^2A^\mu A_
\mu-e^2A_+gA_-g^{-1} -eiA_+g\partial_-g^{-1} -eiA_-g^{-1} 
\partial _+g\right]\quad , \label{gaugedwzwexpl}
\ee 
which represents $\det i\!\!\not\!\!\!\!D$ in terms of bosonic degrees of
freedom. 
Hence we have arrived at a representation of the highly non-trivial
functional $\det i\!\not\!\!\!\! D \lbrack A\rbrack $ in terms of a 
simple effective action, at the expense of introducing as additional 
bosonic group-valued field $g$, over which one has to integrate. As 
seen from (\ref{gaugedwzwexpl}), one recovers
the equivalent bosonic action of free fermions in the limit $e\to 0$, as 
expected. This circumvents the problem of having to replace the 
measure ${\cal D} A$ by ${\cal D}U{\cal D}V$. In this way,
we obtain for the $QCD_2$ partition function
\be
\int {\cal D}A{\cal D}g \E^{i\left( S_{YM} +S_F \lbrack A,g\rbrack \right)}
\label{qcdpartitionf}
\ee
where ${\cal D}A$ stands for the measure and includes the gauge fixing factor.

In the $U(1)$ case, where the Wess-Zumino term in (\ref{wess-zumino-term})
vanishes, only the principal $\sigma$-model is left, that is,
\be 
S=-{1\over 8\pi }\int {\rm d}^2x \, (\partial _\mu \Sigma^{-1})(\partial_\mu 
\Sigma)\quad , \label{wabelian}
\ee  
where $\Sigma  \equiv UV$.
In the $U(1)$ case we consider the parametrization
\be 
U=e^{-i\sqrt\pi (\varphi +\phi)}\quad ,\quad 
V=e^{i\sqrt\pi (\varphi -\phi)}\quad , \quad \Sigma=\E^{-i2\sqrt\pi\phi}
\quad ,\label{u1parametrization}
\ee
leading to
\be
A_+=\sqrt\pi\partial_+(\varphi+\phi)\quad ,\quad
A_-=\sqrt\pi\partial_-(\varphi-\phi)\quad ,\quad
F_{\mu\nu} =\epsilon_{\mu\nu}\sqrt\pi\Box\phi\quad .\label{a-f-for-u1}
\ee
In terms of the $\phi$ field the action reads
\be 
S=\int d^2x\lbrack -{1\over 2} \left(\partial_\mu\phi\right)^2 +
{\pi\over 2e^2}\left(\dal\phi\right)^2\rbrack \quad ,\label{s-u1-of-phi}
\ee
where the last term corresponds to the free action for the gauge field.

Introducing unit in the corresponding partition function, adding to the action
the (quadratic) term
\be 
\delta S=-{e^2\over 2\pi} \int d^2x \left( E+{\pi\over e^2}\dal \phi\right)^2
\ee 
and integrating over the auxiliary field $E$, we find for the
action
\bear S&=&\int d^2x\left[ -{1\over 2}\left(\partial_\mu\phi\right)^2 
-{e^2\over 2 \pi}E^2 +\partial_\mu E\partial^\mu\phi\right]\\ 
&=&\int d^2x\left[ -{1\over 2}\left(\partial_\mu(\phi -E)\right)^2
+{1\over 2}\left(\partial_\mu E\right)^2 -{e^2\over 2\pi}E^2\right]\eear
from which we see that the field $\eta=\phi -E$ is
massless, has negative metric and decouples, realizing the well known
negative metric excitations of the Schwinger model, while $E$ represents the
usual meson of mass $e/\sqrt\pi$. 
\section{Decoupling the dynamics.}

\subsection{QCD$_2$ in the local decoupled formulation and BRST constraints}
The partition function of two-dimensional QCD in the
fermionic formulation (before gauge fixing) is given by the expression
\be
Z=\int{\cal D} A_+{\cal D}A_-\int{\cal D}\psi{\cal D}\bar\psi
e^{iS[A,\psi,\bar\psi]}\label{11qcd2partitionfunction}
\ee
with the action
\be
{\cal S}[A,\psi,\bar\psi]=\int d^2x\left[-\frac{1}{4}\tr F^{\mu\nu}
F_{\mu\nu}+\psi^\dagger_1(i\partial_++eA_+)\psi_1+\psi^\dagger_2(i\partial
_-+eA_-)\psi_2\right].\label{qcd2classaction}
\ee
Our aim is to obtain a bosonic formulation of the theory, in such a way
that structural relations,  hidden in the fermionic formulation are made
clearer in the bosonic counterpart.
We first make some useful change of variables, obtaining a formulation
in terms of matrix-valued fields which decouple at the partition function
level, but which are not totally decoupled, due to the gauge symmetries
of the theory. Later we shall see that there are gauge conditions coupling
the sectors by means of the definition of physical states. In this section,
we deal with the so called local formulation, where the negative metric
field contains physical degrees of freedom, and which is closer to the
usual formulation. Later we will see that one can decouple the negative
metric degrees of freedom in a way very similar to the one used in the 
Schwinger model case, but here the action will have further problems, which
require the introduction of further degrees of freedom.
\subsubsection{Local decoupled partition function and BRST symmetries.}
We can obtain a factorized form of the partition function 
(\ref{11qcd2partitionfunction}) by parametrizing $A_\pm$ as
(\ref{aparametrization}) as well as performing a chiral rotation,
\be 
\psi_1\to\psi_1^{(0)}\equiv U\psi_1,\quad \psi_2\to\psi_2^{(0)}
=V^{-1}\psi_2\quad .\label{ch11transf}
\ee
Denoting the Jacobians by ${\cal J}_G[UV]$ and ${\cal J}_F[UV]$,
and noting that the transformation decouples the fermions, that is,
\bear
\psi_1^\dagger(i\partial_++U^{-1} i\partial_+ U)\psi_1&=&
\psi_1^{(0)\dagger}i\partial_+\psi_1^{(0)},\nonumber\\
\psi_2^\dagger(i\partial_-+Vi\partial_- V^{-1})\psi_2&=&\psi_2^{(0)\dagger}
i\partial_-\psi_2^{(0)},\nonumber
\eear
we arrive at the alternative form of the partition function
(\ref{11qcd2partitionfunction}),
\be
Z=Z_F^{(0)}\int{\cal D}U{\cal D}W{\cal J}_G[W]{\cal J}_F
[W]e^{i {\cal S}_{YM}[W]}\quad ,\label{ch11bosonicpf}
\ee
where  $Z^{(0)}_F$ is the partition function of free fermions
\be
Z^{(0)}_F=\int{\cal D}\psi^{(0)}
{\cal D}\overline\psi^{(0)}\int{\cal D}
e^{i\int d^2x\bar\psi i\slp\psi}\quad ,\label{ch11freeferpf1}
\ee
and ${\cal S}_{YM}$ is the Yang-Mills action
\bear
{\cal S}_{YM}[W]&=&-\frac{1}{4e^2}\int d^2x\tr\frac{1}{2}
[\partial_+(Wi\partial_- W^{-1})]^2 = \int d^2x E^2+E\partial_+ 
(Wi\partial_- W^{-1}) \label{ch11ymaction1}\\
&=&-\frac{1}{4e^2}\int d^2 x\tr\frac{1}{2}[\partial_-
(W^{-1} i\partial_+W)]^2=\int d^2x E^{\prime 2}+E^\prime
\partial_- (W^{-1} i\partial_+W)\quad ,\label{ch11ymaction2}
\eear
with $W=UV$.

Under a vector gauge transformation $U$ and $V$ transform as
$U\stackrel{G}{\rightarrow}
UG$ and $V\stackrel{G}{\rightarrow}G^{-1}V$. Therefore,
$W=UV$ is gauge invariant. In obtaining (\ref{ch11ymaction1}),
(\ref{ch11ymaction2}) for the Yang Mills action we have
first written the field strength tensor $F_{01}$ in terms of
$U$ and $V$ as
\be
F_{01}=-\frac{1}{2}[D_+(U)Vi\partial_- V^{-1}
-\partial_-(U^{-1}i\partial_+U)] =
\frac{1}{2}[D_-(V)U^{-1}i\partial_+U-\partial_+(Vi\partial
_-V^{-1})]\quad ,\label{ch11stresstensor1}
\ee
which can be rewritten in the two alternative forms
\be
F_{01}=-\frac{1}{2}U^{-1}[\partial_+(Wi\partial_- W^{-1})] U=
\frac{1}{2}V[\partial_-(W^{-1}i\partial_+ W)] V^{-1}\quad .
\label{ch11stresstensor2}
\ee

The logarithm of the Jacobian ${\cal J}_F$ is given, following Fujikawa, by
\be
\ln {\cal J}_F=-\ln {\det i\slD \over\det i\not\partial}=-i\Gamma[UV].
\label{ch11jacobifund}
\ee
We can prove the following assertion:
\be
{\cal J}_G[UV]=e^{-ic_V\Gamma[UV]}(\det i\partial_+)_{adj}
(\det i\partial_-)_{adj}\quad ,\label{ch11jacobiadj}
\ee
where $c_V$ is the second Casimir of the group in question with
the normalization $f_{acd}f_{bcd}=c_V\delta_{ab}$
of the structure constants. Representing $(\det i\partial_\pm)_{adj}$
in terms of ghosts and choosing the gauge $U=1$, we obtain
\be
{\cal Z}={\cal Z}^{(0)}_F {\cal Z}^{(0)}_{gh}{\cal Z}
_W\quad ,\label{ch11localpf}
\ee
where ${\cal Z}^{(0)}_{gh}={\cal Z}^{(0)}_{gh+}{\cal Z}^{(0)}_{gh-}$
and
\be
{\cal Z}_W=\int{\cal D}W e^{-i(1+c_V)\Gamma[W]}
e^{i{\cal S}_{YM}[W]}=\int{\cal D}W e^{-iS_{eff}[W]}\quad .\label{ch11localwpf}
\ee
In the above expression the effective action is given by
\be
S_{eff}[W] = S_{YM} -(c_V +1)\Gamma [W]\label{localeffecaction}
\ee
and has two contributions, one corresponding to a WZW action with a 
negative coefficient, and the Yang-Mills action written in either of the 
forms (\ref{ch11ymaction1}) or (\ref{ch11ymaction2}).

We refer to (\ref{ch11localpf}) as the ``local decoupled''
partition function. As seen from (\ref{ch11localwpf}), the ghosts
$b_\pm^{(0)}$ are canonically conjugate to $c_\pm^{(0)}$ and have 
Grassmann parity $+1$. We assign to them the ghost number 
$gh\# = -1$ and $gh\# = +1$, respectively.

The dimensionality of the direct-product space
${\cal H}_F^{(0)}\otimes {\cal H}_{gh}^{(0)}\otimes {\cal H}_W$
associated with the partition function (\ref{ch11localpf}) is
larger than that of the physical Hilbert space of the original
fermionic formulation. Hence there must exist constrains imposing restrictions
on the representations which are allowed in ${\cal H}_{phys}$. In order
to discover these constraints we observe that the partition function
is separately invariant under the following nilpotent transformations\\
\pspace\parbox{6cm}{\bearst
&&W\delta W^{-1}=- c_-^{(0)}\; , \\
&&\delta\psi^{(0)}_1= c_-^{(0)}\psi_1^{(0)}\; ,\;
\delta\psi^{(0)}_2=0\; ,\\
&&\delta c_-^{(0)}=\frac{1}{2}\{c_-^{(0)} , c_-
^{(0)}\} \; , \;\delta c_+^{(0)}=0\; ,\\
&&\delta b^{(0)}_- =\Omega_-
\; , \;
\delta b_+^{(0)}=0\quad ,\eearst}
\hfill
\parbox{6cm}{\bearst
&&W^{-1}\delta W=- c_+^{(0)}\; ,\\
&&\delta\psi_1^{(0)}=0\; ,\; \delta\psi_2^{(0)}=
c_+^{(0)}\psi_2^{(0)}\; ,\\
&&\delta c^{(0)}_-=0\; ,\; \delta c_+^{(0)}=\frac{1}{2}
\left\{c_+^{(0)},c_+^{(0)}\right\}\; ,\\
&&\delta b_-^{(0)}=0\; ,\; \delta b_+^{(0)}=\Omega_+\; ,
\eearst}\hfill\peight{\bear\label{localbrstplus}\eear}\\
where $\delta$ denotes the variation graded with respect to Grassmann
parity, and 
$\Omega_\mp$ are given by\\
\peffec{\bearst
\Omega_- &=& -\frac{1}{4e^2}{\cal D}_-(W)(\partial_+(Wi\partial_-
W^{-1}))-\left({1+c_V}\right)
J_- (W) +j_- \\
\Omega_+&=&-\frac{1}{4e^2}{{\cal D}}_+(W)(\partial_-(W^{-1}i\partial_+
W))-\left(1+c_V\right)
J_+ (W)+j_+
\eearst}\hfill\peight{\bear\label{omegaminusplus}\eear}\\
with
\bear
&&J_- (W)= {1\over 4\pi}Wi\partial_- W^{-1}\quad ,\quad
J_+ (W) = {1\over 4\pi} W^{-1}i\partial_+W\label{jwzwpm}\\
&&j_-=\psi_1^{(0)}\psi_1^{(0)\dagger}+\{b_-^{(0)},c_-^{(0)}\}
\quad , \quad
j_+=\psi_2^{(0)}\psi_2^{(0)\dagger}+\{b_+^{(0)},c_+^{(0)}\}\quad .
\label{jzeropm}\eear
These transformations are easily derived by departing from two 
alternative forms (\ref{ch11ymaction1}) or (\ref{ch11ymaction2}) of the 
Yang-Mills action.

The corresponding BRST currents,
as obtained via the usual Noether construction, are found to be
\be
J_\mp  =\tr c^{(0)}_-\left[\Omega_\mp 
-\frac{1}{2}\{b_\mp^{(0)},c_\mp^{(0)}\}\right]\label{noetherbrstmp}
\ee
with $\partial_+J_- =0$, and $\partial_-J_+ =0$.

Remarkably enough, the nilpotent symmetries lead to  currents $J_-$,
$J_+$ which only depend on the variable $x^-$ and $x^+$, respectively.

The on-shell nilpotency of the corresponding conserved charges
\be
Q_\pm =\int dx^1 J_\pm (x^\pm)\label{correspconscharges}
\ee
follows from the first-class character of the operators $\Omega^a_\pm
=\tr \left( t^a\Omega_\pm\right)$.
\subsection{$QCD_2$ in the non-local decoupled formulation and BRST 
constraints}
The partition function represented by the standard expression
(\ref{ch11localwpf}) contains fields which are mixtures of massive 
and massless modes, of positive and negative norms respectively, 
coupled by the constraints. In the following section we dissociate
these degrees of freedom by a suitable 
transformation. We shall thereby be lead to an alternative
nonlocal representation of the partition function, useful for learning 
certain structural properties.
\subsubsection{Non-local decoupled partition function and BRST symmetries}
Following ref.  \cite{aaijmpa}, we  make in (\ref{ch11ymaction1}) the 
change of variable $E\to\beta$ defined by
\be
\partial_+E = \left(\frac{1+c_V}{2\pi}\right)\beta^{-1}i\partial_+\beta\quad .
\label{eofbeta}
\ee
The Jacobian  associated with this change of variables
is
\be
{{\cal D}}E =  
\det i{\cal D}_+ (\beta) {{\cal D}}\beta\label{deofdbeta}
\ee
where we have suppressed the constant $\det \partial_+$ which will not play
any role in the discussion to follow. Making use of the determinant of the
fermionic operator in the adjoint representation (which corresponds
to the determinant in the fundamental representation raised to the
power $c_V$) as well as (\ref{ch11jacobifund}), and representing 
$\left( \det i\partial_+\right)$ as a functional integral over ghost 
fields $\hat b_-$ and $\hat c_-$, we have, after decoupling the ghosts,
\bear
Z&=&Z_F^{(0)}Z_{gh}^{(0)} \hat Z_{gh-}^{(0)}\int{\cal D}W\int{\cal D}
\beta \exp\{-i(1+c_V)[\Gamma[W]+\Gamma[\beta]
\nonumber\\
& & -\frac{1}{4\pi}\int tr(\beta^{-1}
\partial_+\beta W\partial_-W^{-1})]\} \label{ch11nonlocalpf}\\
& & \times \exp({ i\Gamma[\beta]})
\exp\left\{{i\left(\frac{1+c_V}{2\pi}\right)^2
e^2\int\frac{1}{2}tr\left[\partial^{-1}_+(\beta^{-1}\partial_+\beta
\right]^2}\right\}\nonumber
\eear
where 
\be
\hat Z_{gh-}^{(0)} = \int{\cal D}\hat b^{(0)}_-{\cal D}\hat 
c_-^{(0)} e^{i\int d^2 x\tr\hat 
b_-^{(0)}i\partial_+\hat c^{(0)}_-}\label{ghosthatminuspf}
\ee
Using the Polyakov-Wiegmann identity (\ref{pol-wieg-id})
and making the change of variable $W\to\beta W=\tilde W$,
we are left with
\be
Z=Z^{(0)}_FZ^{(0)}_{gh} \hat Z_{gh-}^{(0)}
Z_{\tilde W}Z_\beta\label{z=prodzsnonlocal}
\ee
where
\be
Z_\beta=\int{{\cal D}}\beta \exp\left\{{i\Gamma[\beta]
+i\left(\frac{1+c_V}{2\pi}
\right)^2e^2\int \frac{1}{2}tr\left[\partial_+^{-1}
(\beta^{-1}\partial_+\beta)\right]^2}\right\}\label{zbeta}
\ee
and
\be
Z_{\tilde W}=\int{\cal D}\tilde W \exp[{-i(1+c_V)\Gamma[\tilde W]}]
\label{negativemetricfieldpf}
\ee
is the partition function of a WZW field of level $-(1+c_V)$.
Note that the decoupling of the $\beta$-field depended crucially
on the choice of the multiplicative constant in (\ref{eofbeta}).
Repeating this process by starting from expression (\ref{ch11ymaction2})
and making now the change of variables
\be
\partial_+E^\prime =
\left(\frac{1+c_V}{2\pi}\right)\beta^\prime i\partial_-\beta^{\prime -1}
\label{eprimeofbetaprime}
\ee
and
\be
W\to W\beta^\prime =\tilde W^\prime\quad .\label{wtowprime}
\ee
one arrives at the equivalent representation
\be
Z=Z^{(0)}_FZ^{(0)}_{gh} \hat Z_{gh+}^{(0)}
Z_{\tilde W^\prime}Z_{\beta^\prime}\label{z=prodzsnonlocalprime}
\ee
where
\bear
\hat Z_{gh+}^{(0)} &=& \int{\cal D}\hat b^{(0)}_+{\cal D}\hat 
c_+^{(0)} e^{i\int d^2 x\tr\hat b_+^{(0)}i\partial_-\hat
c^{(0)}_+}\quad ,\label{zghplus}\\
Z_{\beta^\prime}&=&\int{\cal D}\beta^\prime \exp\left\{{i\Gamma[\beta^\prime]
+i\left(\frac{1+c_V}{2\pi}
\right)^2e^2\int \frac{1}{2}tr\left[\partial_+^{-1}
(\beta^\prime\partial_+\beta^{\prime -1})\right]^2}\right\}\quad ,
\label{zbetaprime}\\
Z_{\tilde W^\prime}&=&\int{{\cal D}}\tilde W^\prime 
\exp[{-i(1+c_V)\Gamma[\tilde V]}]\quad .
\label{zvprime}
\eear

The partition function (\ref{z=prodzsnonlocalprime}) exhibits
nilpotent symmetries in a variety of sectors, not all of which are to be
imposed as symmetries of the physical states. In the following section we 
systematically derive those nilpotent symmetries which have to be
realized in order that the theory is consistently defined and represents
two-dimensional QCD.

\subsubsection{Systematic derivation of the constraints}
The main objective of this section is to trace the fate of the BRST
conditions of the local formulation, when going over to
the so-called\cite{aaijmpa} non-local formulation. We shall also establish
from first principles further  BRST conditions that may have to be imposed 
in order to ensure the equivalence of local and non-local formulations.
\subsubsection{\it a) The BRST condition $Q _+\approx 0$ in the non-local 
formulation}
Making use of some algebraic identities, we may rewrite $\Omega_+$ in 
(\ref{omegaminusplus}) as
\be
\Omega_+ = -\frac{1}{4e^2}W^{-1}\left[\partial_+^2(Wi\partial_-
W^{-1})\right]W-(1+c_V)J_+(W)+j_+.\label{omegaplusrewritten}
\ee
Using the equation of motion for $E$, $\Omega_+$ and making the change of 
variable (\ref{eofbeta}), we then obtain
\be
\Omega_+=-(1+c_V)J_+(\tilde W)+j_+, \label{omegaplusrewritten3}
\ee
where $\tilde W=\beta W$. 
We conclude that the corresponding nilpotent charge
\be
Q_+=\int dx^1tr c_+^{(0)}\left[-(1+c_V)J_+(\tilde W)
+j_+-\frac{1}{2}\left\{b_+^{(0)},c_+^{(0)}\right\}\right],\label{nilpocharge}
\ee
must annihilate the physical states.

\subsubsection{\it b) The BRST condition $Q_-\approx 0$ in the non-local
formulation}
In the case of the BRST charge $Q_+$, the symmetry transformations
in the $V$-fermion-ghost space giving rise to this conserved
charge could be trivially extended to the $E-W$-fermion-ghost
space. This is not true for $Q_-$, where
the BRST symmetry for $E$ is maintained off-shell only
by adding  an extra (commutator) term (which vanishes for
$E$ ``on shell''). One is thereby led to a fairly complicated expression
for $Q_-$ in terms of the variables $\beta, \tilde
W$ of the non-local formulation.

A more transparent result is obtained by rewriting $\Omega_-$  
in terms of the variables $\tilde W^\prime$ defined
in (\ref{wtowprime}). Making use of the $E^\prime$ equation of motion
and making the corresponding change of variables (\ref{eprimeofbetaprime}),
we may rewrite $\Omega_-$ in (\ref{omegaminusplus}) in the form
\be
\Omega_-=-(1+c_V)J_-(\tilde W')+j_-.\label{omegaminusrewritten2}
\ee
We conclude that the corresponding BRST charge
\be
Q_-=\int dx^1tr c_-^{(0)}[-(1+c_V)J_-(\tilde W')
+j_-\frac{1}{2}\{b_-^{(0)},c_-^{(0)}\}]\label{qminusbrst}
\ee
must annihilate the physical states.

Although $\tilde W$ in (\ref{nilpocharge}) and $\tilde W'$ in 
(\ref{qminusbrst}) obey the same dynamics, as described by the WZW
action, they are related by different
constraints to the ``massive'' sector, which is described in terms
of the group-valued fields $\beta$ and $\beta'$, respectively,
obeying different dynamics. We will return to this question
in the next section.

\subsubsection{\it c) BRST condition associated with the changes of variables
$E\to\beta$ and $E^\prime\to\beta^\prime$.}
The change of variable (\ref{eofbeta}) leads to
a BRST condition on the physical states. We follow the procedure
outlined in ref. \cite{bastianelli}, arriving at the constraint
$\tilde\Omega_-$, given by
\bear
{\hat\Omega}_-&=&-\lambda^2\beta(\partial_+^{-2}(\beta^{-1}
i\partial_+\beta))\beta^{-1}+J_-(\beta)\nonumber\\
&&-(1+c_V)J_-(\tilde W)+\left\{{\hat b}_-^{(0)},\hat c_-^{(0)}
\right\}.\label{tildeomegaminus}
\eear
The corresponding Noether current is found to be
\be
\hat J_-=\tr\hat c_-^{(0)}\left[\hat\Omega_- -\frac{1}{2} \left\{b_-^{(0)}
\hat c_-^{(0)}\right\}\right].\label{jhatminusofcomega}
\ee
with $\partial_+\hat J_- =0$.
Our deductive procedure shows that the corresponding
nilpotent charge $\hat Q_-$ must annihilate the physical
states:
\be
\hat Q_-=0\quad {\rm on}\quad {\cal H}_{phys}.\label{qiszeroonhphys}
\ee

Repeating this procedure for the change of variable
(\ref{eprimeofbetaprime}) in (\ref{ch11ymaction2}), one is led to conclude
that the left-moving current
\be
\hat J_+=\tr\hat c_+^{(0)}\left[\hat\Omega_+ -\frac{1}{2} \left\{b_+^{(0)}
\hat c_+ ^{(0)}\right\}\right]\quad ,\label{jhatplusofcomega}
\ee
which obeys $\partial_- \hat J_+ =0$, and where
\bear
\hat\Omega_+&=&-\lambda^2{\beta'}^{-1}(\partial_-^{-2}
(\beta' i\partial_-{\beta'}^{-1}))\beta'+J_+(\beta')\nonumber\\
&&-(1+c_V)J_+(\tilde V')+\{\hat b_+^{(0)},\hat c_+^{(0)}\}.
\label{omegahatplus}
\eear
must also annihilate the physical states. On the zero-ghost-number sector
conditions $\hat Q_\pm\approx 0$ are again equivalent to requiring
$\hat\Omega_\pm\approx 0$.
In summary, on the zero-ghost-number sector, the BRST conditions which should
be imposed on the physical states are equivalent to requiring
\be
\Omega_\pm\approx0,\quad\hat\Omega_\pm\approx0,\label{omegaomegahat=0}
\ee
with $\hat\Omega_\pm$ and $\Omega_\pm$ given by
eqs. (\ref{omegaplusrewritten3}), (\ref{omegaminusrewritten2}), 
(\ref{tildeomegaminus}), and (\ref{omegahatplus}). It is interesting
to observe that the gauging procedure of \cite{kara-sch} would suggest the
existence of further constraints. It thus merely provides a guideline
for discovering candidates to constraints.
\subsection{The physical Hilbert space}
In order to address the cohomology problem defining the
physical Hilbert space, we must express the constraints
in terms of canonically conjugate variables.
We first rewrite the partition function $Z_\beta$ in (\ref{zbeta})
in terms of an auxiliary field $B$ as
\be
Z_\beta=\int {{\cal D}}B{{\cal D}}\beta e^{iS[\beta,B]},\label{zbetaphyshil}
\ee
where
\be
S[\beta,B]=\Gamma[\beta]+\int tr\left[\frac{1}{2}
(\partial_+B)^2+\lambda B\beta^{-1}i\partial_+\beta\right].
\label{sbetaphyshil}
\ee
Similarly, we have for $Z_{\beta'}$ in (\ref{zbetaprime})
\be
Z_{\beta'}=\int {{\cal D}}B'{{\cal D}}\beta' e^{iS'[\beta',B']}, 
\label{zbetaprimephyshil}
\ee
with
\be
S'[\beta',B']=\Gamma[\beta']+\int tr\left[\frac{1}{2}
(\partial_-B')^2+\lambda B'\beta'i\partial_-{\beta'}^{-1}\right].
\label{sbetaprimephyshil}
\ee
We may then rewrite the constraints ${\hat\Omega}_\pm\approx 0$
in (\ref{tildeomegaminus}) and (\ref{omegahatplus}) as
\bear
\hat\Omega_-&=&\lambda\beta B\beta^{-1}+\frac{1}{4\pi}\beta i
\partial_-\beta^{-1}-\frac{(1+c_V)}{4\pi}\tilde V i\partial_-
\tilde V^{-1}+\{\hat b_-^{(0)},\hat c_-^{(0)}\},
\label{omegahatminusrewrittennew}\\
\hat\Omega_+&=&\lambda{\beta'}^{-1}
 B'\beta'+\frac{1}{4\pi}{\beta'}^{-1}
i\partial_+\beta'-\frac{(1+c_V)}{4\pi}\tilde {V'}^{-1} i\partial_+
\tilde V'+\{\hat b_+^{(0)},\hat c_+^{(0)}\}.
\label{omegahatplusrewrittennew}
\eear
Canonical quantization then allows one to obtain the Poisson 
algebra. It is straightforward to verify that $\hat\Omega^a_+=tr(\hat\Omega 
t^a)$ and $\hat{\Omega}^a_-=tr(\hat\Omega t^a)$ are first class:
\be
\left\{\hat\Omega_\pm^a(x),\hat\Omega^b_\pm(y)\right\}_{PB}=-f_{abc}
\hat\Omega^c_\pm\delta(x^1-y^1). \label{omegaisfristclass}
\ee
Hence the corresponding BRST charges are nilpotent. Similar
properties are readily established for the remaining operators
$\Omega_\pm$. Furthermore,
\be
\lbrace\Omega_+(x), \hat\Omega_-(y)\rbrace_{PB}=0\quad , \quad
\lbrace\Omega_-(x), \hat\Omega_+(y)\rbrace_{PB}=0\quad .
\label{omegawithomegahat}
\ee

The physical Hilbert space of the non-local formulation of
$QCD_2$ is now obtained by solving the cohomology problem
associated with the BRST charges $Q_\pm,\hat Q_\pm$ in the
zero-ghost-number sector. The solution of this problem is
suggested by identifying this space with the space of
gauge-invariant observables of the original theory. It is 
interesting to note that the constraints $\hat\Omega_\pm\approx 
0$ are implemented by any functional of $V$ (and the fermions), 
thus implying that $\tilde V,\beta(\tilde V', \beta')$ can only occur
in the combinations $\beta^{-1}\tilde V(\tilde V'{\beta'}^{-1})$.
Indeed, making use of the Poisson brackets defined by the canonical
procedure, we have\\
\ptwelve{\bearst
\left\{\hat\Omega_-^a(x),\ \beta^{-1}(y)\right\}_{PB}&=&
i(\beta^{-1}(x)t^a)\delta(x^1-y^1), \\
\left\{\hat\Omega_-^a(x),\ \tilde V(y)\right\}_{PB}&=&
-i(t^a\tilde V(y))\delta(x^1-y^1), \\
\left\{\hat\Omega_+^a(x),\ \tilde V'(y)\right\}_{PB}&=&
+i(\tilde V'(x)t^a)\delta(x^1-y^1), \\
\left\{\hat\Omega_+^a(x),\ {\beta'}^{-1}(y)\right\}_{PB}&=&
-i(t^a{\beta'}^{-1}(y))\delta(x^1-y^1)\quad . \eearst}\hfill
\peight{\bear\label{omegahatspb}\eear}\\
As for the other two constraints, $\Omega_+\approx 0$
and $\Omega_-\approx0$, which link the bosonic to the free
fermionic sector, they indicate that local fermionic
bilinears should be constructed in terms of free fermions and the
bosonic fields as
\be
\left(\psi_1^{(0)\dagger}\beta^{-1}\tilde V\psi_2^{(0)}\right)
=\left(\psi_1^{(0)\dagger}\tilde V'{\beta'}^{-1}\psi_2^{(0)}\right)
=\left(\psi_1^{(0)\dagger}V\psi_2^{(0)}\right)
=\left(\psi_1^\dagger\psi_2\right).\label{physhilcompo}
\ee
This is in agreement with our expectations.
\section{Massive two-dimensional QCD}
In this section we shall see that the BRST symmetries of the physical states 
in massless QCD$_2$ are also the symmetries which should be imposed on the 
physical states in the massive case.

For massive fermions the functional determinant of the 
Dirac operator, an essential ingredient for arriving at the bosonised form 
of the QCD$_2$ partition function, can no longer be computed in 
closed form, and one must resort to the so-called adiabatic principle of 
form invariance. Equivalently, one can start with a perturbative expansion 
in powers of the mass, as given by
\be
\sum {1\over n!}M^n \left\lbrack  \int {\rm d}^2 x\overline \psi \psi\right\rbrack ^n\quad ,
\ee
use the (massless) bosonization formulae and re-exponentiate the
result. In this approach, the mass term is given in terms of a 
bosonic field $g_\psi $ of the massless theory by\cite{gepner}
\bearst
S_m =- M\int \overline \psi\psi = M\mu\int \tr (g_\psi + g_\psi^{-1})\quad ,
\eearst
where $\mu$ is an arbitrary massive parameter whose value depends  on the
renormalization  prescription for the mass operator.\cite{aar}

Defining $m^2 = M\mu$, we re-exponentiate the mass term. Going through the 
changes of variables leading to (\ref{z=prodzsnonlocal}) and 
(\ref{z=prodzsnonlocalprime}), one arrives at the following 
alternative forms for the mass term when expressed in terms of the fields 
of the non-local formulation
\bear
S_m &=& m^2 \int\tr (g\tilde\Sigma^{-1} \beta + \beta^{-1} \tilde
\Sigma g^{-1})\quad ,
\label{2.3}\\
S_m ^\prime &=&  m^2 
\int\tr (g\beta^\prime\tilde \Sigma^{^\prime -1} + 
\tilde\Sigma^\prime\beta^{\prime -1}  g^{-1})\quad . \label{2.3prime}
\eear

The corresponding effective action of massive QCD$_2$ in the non-local 
formulation reads
\bear
S&=&S_{YM}\lbrack \beta, B\rbrack +S_m\lbrack g,\beta,\tilde\Sigma\rbrack +
\Gamma\lbrack g\rbrack  + \Gamma \lbrack \beta\rbrack  
- (c_V+1)\Gamma\lbrack \tilde\Sigma\rbrack +S_{gh}+
 \hat S_{gh-}\quad , \label{massive-boson-action}\\
S^\prime &=&S_{YM}^\prime \lbrack \beta^\prime , B^\prime \rbrack +
S_m\lbrack g,\beta^\prime ,\tilde\Sigma^\prime \rbrack +
\Gamma\lbrack g\rbrack  + \Gamma \lbrack \beta^\prime \rbrack  
- (c_V+1)\Gamma\lbrack \tilde\Sigma^\prime \rbrack +
S_{gh} +\hat S_{gh+} \quad . \label{massive-boson-action-prime}
\eear

We thus see that the associated partition function no longer 
factorizes. Nevertheless, there still exist BRST currents which are either 
right- or left-moving, just as in the massless case.

The actions (\ref{massive-boson-action}) 
and (\ref{massive-boson-action-prime}) 
exhibit various symmetries of the BRST type;  however, not all of 
them lead to nilpotent charges. The variations are graded with 
respect to Grassmann number. The equations of motion obtained 
from action (\ref{massive-boson-action}) read
\bear
{1\over 4\pi}\partial_+(g\partial_-g^{-1}) =\, &&  m^2 (g\tilde
\Sigma^{-1}\beta -
\beta^{-1} \tilde\Sigma g^{-1})\quad ,\label{massiveeqg}\\
-{c_V+1\over 4\pi}\partial_+(\tilde\Sigma \partial_-\tilde\Sigma^{-1}) =
\, && m^2 (\tilde\Sigma g^
{-1}\beta^{-1} - \beta g\tilde \Sigma^{-1})\quad, \label{massiveeqtildesigma}\\
{1\over 4\pi} \partial_+ (\beta \partial_-\beta^{-1}) +
i \lambda\partial _+(\beta B
\beta^{-1}) =
\, && m^2 (\beta g \tilde\Sigma^{-1} - \tilde
\Sigma g^{-1}\beta^{-1})\quad, \label{massiveeqbeta}\\
-{1\over 4\pi}\partial_-(\beta^{-1}\partial_+\beta)+i\lambda\lbrack
\beta^{-1} \partial_+ \beta, B\rbrack  + && \nonumber\\
i \lambda \partial_+B = \, && m^2 (g\tilde\Sigma^{-1}\beta -
\beta ^{-1}\tilde\Sigma g^{-1})\quad ,\label{massiveeqbeta2}\\
\partial_+^2 B = \, && \lambda (\beta ^{-1}i\partial_+\beta)\quad ,
\label{massiveeqb}\\
\partial_\pm b_\mp = \, && 0 \quad ,
\quad \partial _\pm c_\mp =0 \quad ,\label{massiveeqgh}
\eear
with an analogous set of equations involving the prime sector.
Notice that the mass term can be transformed from one equation 
to another, by a suitable conjugation. Making use of eqs. 
(\ref{massiveeqg}-\ref{massiveeqgh}), 
the Noether currents are constructed in the standard fashion: we make a 
general BRST variation of the action, without  using the equations of 
motion, and equate the result to the on-shell variation, taking into 
account terms arising from partial integrations. The only subtlety in 
this procedure concerns the WZW term, which only contributes off-shell to 
the variation. The four conserved Noether currents are found to be
\bear
J_\pm =\, && \tr \left( c_\pm\Omega_+ -
{1\over 2} b_\pm\{c_\pm, c_\pm\}\right)
\quad ,\label{2.10a}\\
\hat J_\pm=\, && \tr \left( c_\pm \hat\Omega_\pm -
{1\over 2}\hat  b_\pm \{\hat c_\pm, \hat c_\pm\}\right)\quad ,\label{2.10d}
\eear
where the $\Omega $'s are given by
\bear
\Omega_+  &=& \left({1\over 4\pi}g^{-1}i\partial _+g -
{c_V +1\over 4\pi}
\tilde\Sigma^{-1}i\partial_+\tilde\Sigma + \{b_+, c_+\}\right)\quad,
\label{2.8a}\\
\Omega_-  &=& \left( {1\over 4\pi}g i\partial _- g^{-1} -
{c_V +1\over 4\pi}\tilde\Sigma^\prime i\partial_-\tilde\Sigma^{\prime -1} +
\{b_-, c_-\}  \right)\quad,
\label{2.8b}\\
\hat\Omega_- &=&   \left( {1\over 4\pi}\beta i\partial _- \beta^{-1} -
 {c_V+1\over 4\pi}\tilde \Sigma i\partial_-\tilde\Sigma^{-1} -
 \lambda \beta B \beta^{-1} +
\{b_-, c_-\}\right) \quad , \label{2.8c}\\
\hat\Omega_+ &=&  \left( {1\over 4\pi}\beta^{\prime -1} i\partial _+
\beta^\prime - {c_V+1\over 4\pi}\tilde \Sigma^{\prime -1} i\partial_+
\tilde\Sigma^\prime - \lambda \beta^{\prime -1} B^\prime \beta^\prime +
\{\hat b_+,\hat c_+\}\right) \quad . \label{2.8d}
\eear
From the current conservation laws
\be
\partial_\mp J_\pm=0 \quad ,\quad \partial _\pm\hat J_\mp =0 
\quad ,\label{2.9}
\ee
one infers that $\Omega_-$ and $\hat \Omega_-$ 
are right-moving, while $\Omega_+$ and $\hat\Omega_+$ are left-moving. 
Indeed, making use of the equations of motion 
(\ref{massiveeqg}-\ref{massiveeqgh}) 
one readily checks that the operators 
$\Omega_\pm$, $\hat \Omega_\pm$ satisfy
\be
\partial_\mp\Omega_\pm=0 \quad ,\quad \partial _\pm\hat\Omega \mp=0 
\quad ,\label{2.12}
\ee
consistent with the conservation laws (\ref{2.9}).

\section{Screening in two-dimensional QCD}
In this section we reconsider the problem of screening and confinement.
We shall concentrate on the case of single flavour QCD, and merely comment 
on the general case at the end of the section.

We proceed by first considering the case of massless fermions and compute
the inter-quark potential. We introduce a pair of classical colour 
charges of strength $q=q^at^a$ separated by a distance $L$. Such a pair
is introduced in the action (\ref{massive-boson-action}) by means of 
the substitution
\be
i(\beta^{-1}\partial_+\beta)^a\longrightarrow 
i(\beta^{-1}\partial_+\beta)^a-{2\pi\over e}q^a\bigg(\delta(x-{L\over 
2})-\delta(x+{L\over 2})\bigg)\quad ,\label{11externalchargesubstitution}
\ee
where $a$ is a definite colour index. This adds the following term to 
the action 
\footnote{This corresponds to minus the same term added to the hamiltonian.}
\be
V(L)=\Delta S=S_q-S=-(c_V+1)
q^a\bigg(B^a(L/2)-B^a(-L/2)\bigg)\quad .\label{11potential-masslesscase} 
\ee
The equation of motion for $B^a$ is now replaced by
\be
\partial_+^2B^a=i\lambda(\beta^{-1}\partial_+\beta)^a
-(c_V+1)q^a\bigg(\delta(x-{L\over 2})-\delta(x+{L\over 2})\bigg),
\ee
which implies, upon substitution into the equation of motion for the 
$\beta$-field,
\begin{eqnarray}
& &\partial_+\left({i\over 4\pi\lambda}\partial_-\partial_+ B+[\partial_+ 
B,B]+i\lambda B\right)=\nonumber\\
& &\qquad\left({-iq\over 2e}\partial_-
+(c_V+1)[q,B]\right)
\lbrack\delta(x-{L\over 2})-\delta(x+{L\over 
2})\rbrack\quad .\label{11massivebetaeqwithextch}
\end{eqnarray}
We look for solutions of (\ref{11massivebetaeqwithextch}) with a fixed global 
orientation in colour space.\footnote{Note that this is a non-trivial
input, since we have no longer the freedom of choosing a gauge in which
such an Ansatz could be realized}  We thus make  $B^a=q^a f(x)$. This
renders the problem abelian. We thus infer that the potential 
(\ref{11potential-masslesscase}) has the form
\be
V(L)= {(c_v+1)\sqrt\pi\over 2}{q^2\over e}(1-e^{-2\sqrt\pi\lambda L})
\label{11potential-masslesscasefin}
\ee
which implies that the system is in a screening phase.

We now turn to the case of massive fermions. Taking the external charge to lie
in the direction $t^2$ of $SU(n)$ space, our Ansatz for $B^a$ leads one to look
for solutions with $g$, $\beta$ and $\Sigma$ parametrized as
\be
g=e^{i2\sqrt\pi\varphi\sigma_2},\qquad \beta=e^{i2\sqrt\pi 
E\sigma_2},\qquad \Sigma=e^{-i2\sqrt\pi\eta\sigma_2}\quad ,
\label{gbetawparametrizationt2}
\ee
The equations of motion (\ref{massiveeqg}-\ref{massiveeqb})
are replaced by\footnote{We leave the 
Casimir $c_V$ as a free parameter, since the
expressions corresponding to the Schwinger model will simply be 
obtained from the SU(N) model by taking the limit $c_V\to 0$.}
\begin{eqnarray}
\partial_+\partial_-\varphi&=&-4\sqrt\pi m^2 {\rm 
sin}2\sqrt\pi(E+\varphi+\eta)\quad ,\label{11massiveeqvarphi}\\
\partial_+\partial_-\eta&=&{4\sqrt\pi\over c_v+1}m^2 {\rm 
sin}2\sqrt\pi(E+\varphi+\eta),\label{11massiveeqeta}\\ 
\partial_+\partial_- E+4\pi\lambda^2 E&=&-4\sqrt\pi m^2{\rm 
sin}2\sqrt\pi(E+\varphi+\eta)\nonumber\\
& &-2\sqrt\pi(c_v+ 1)\lambda 
q\!\!\left[\Theta(x+{L\over 2})-\Theta(x-{L\over 2})\right]\quad .
\label{11massiveeqe}
\end{eqnarray}

We notice the  combination $\psi =\varphi+(c_v+1)\eta$
describes a zero-mass field. In order to compute the potential
we consider the static limit. This is equivalent to considering the effective 
lagrangean
\begin{eqnarray}
{\cal L}&=&{1\over 2}{\partial_1E}^2+{1\over 2}{c_V+1\over 
c_V}{\partial_1\Phi}^2+2\pi\lambda^2 E^2-2m^2{\rm 
cos}2\sqrt\pi(\Phi+E)\nonumber\\
& &+2\sqrt\pi(c_V+1)\lambda q E\left[\Theta(x+{L\over 2})-\Theta(x-{L\over 
2})\right]\label{11effeclagparafields}\\
& &-{(c_V+1)^2 q^2\over 2}\left[\Theta(x+{L\over 
2})-\Theta(x-{L\over 2})\right]-{{\partial_1\psi}^2\over 2 c_v}\quad ,\nonumber
\end{eqnarray}
where $\Phi=\varphi+\eta$. In order to compute the inter-quark potential,
we shall expand the cosine in the effective Lagrangian
(\ref{11effeclagparafields}) up to second order in the argument. This
pre-supposes a bound in the fluctuations of the fields.
We can confirm that the solution is consistent with such a condition.

In the weak-limit approximation, we expand the cosine term. 
Consequently, we diagonalise the hamiltonian and solve 
the equations of motion.\footnote{All the forthcoming computations will 
in general be valid for any compact group. In such cases, the mass term 
can always be expanded in terms of
algebra-valued fields after a convenient parametrisation.}
The diagonalisation of the quadratic lagrangean leads to 
the expression
\begin{eqnarray}
{\cal L}&=&-{1\over 2 c_V}{\partial_1\psi}^2 
+(1+\epsilon a^2)\left\{{1\over 2}{\chi^\prime}^2_++{1\over 
2}m^2_+\chi_+^2+\lambda Q_+\chi_+\right\}\nonumber\\
&+&{(1+\epsilon a^2)\over a^2}\left\{{1\over 2}{\partial_1\chi}^2_-+{1\over 
2}m_-^2\chi_-^2+\lambda Q_-\chi_-\right\}\quad ,\label{11effeclagparafieldschi}
\end{eqnarray}
where we have found it useful to define the following variables :
\begin{eqnarray}
\chi_+={1\over 1+\epsilon a^2}(E-a\Phi)\quad ,\quad
\chi_-={1\over 1+\epsilon a^2}(\Phi+\epsilon a E)
\end{eqnarray}
and the parameters:\\
\peffec{\bearst
\epsilon&=&{c_V\over (c_v+1)}\quad ,\quad  q_+={2\sqrt\pi(c_v+1)q\over 
(1+\epsilon a^2)}, \\
a&=&-{8\pi  m^2\over m_+^2- 16\epsilon  m^2}\quad ,\quad
q_-={2\sqrt\pi\epsilon a (c_v+1) q\over (1+\epsilon
a^2)}\\
m_{\pm}^2&=&2\pi\lbrack\left(\lambda^2+(1+\epsilon)2  m^2\right)
\pm\sqrt{\left(\lambda^2+(1+\epsilon)2m^2\right)^2-8\epsilon\lambda^2  
m^2}\rbrack,\nonumber\\
Q_\pm&=&q_\pm \left[\Theta(x-{L\over 2})-\Theta(x+{L\over 2})\right]\quad .
\eearst}\hfill\peight{\bear\label{11allparameters}\eear}\\
Solving the corresponding equations of motion yields:
\be
\chi_\pm={\lambda q_\pm\over m_\pm^2} 
\left\{\begin{array}{ll}
{\rm sinh}(m_\pm {L\over 2})e^{-m_\pm|x|}  &  \qquad |x|>{L\over 2}
\\
(1-e^{-m_\pm L/2}{\rm cosh} m_\pm x)  &  \qquad |x|<{L\over 2}
\end{array}
\right.\label{chiascases}
\ee
\noindent
from which we obtain the inter-quark potential energy
\begin{eqnarray}
V(L)=&&{(c_V+1)^2q^2\over 2}\nonumber\\
& &\quad\times\left[ 
\left({4\pi\lambda^2-m_-^2\over 
m_+^2-m_-^2}\right)\!\!\!\left({1-e^{-m_+L}\over m_+}\right)
+\left({m_+^2-4\pi\lambda^2\over 
m_+^2-m_-^2}\right)\left({1-e^{-m_-L}\over 
m_-}\right)\right]\; .\label{interquarkpotential}
\end{eqnarray}
Thus we find two mass scales given by $m_+$ and $m_-$. Both these 
scales correspond to screening-type contributions if $c_V\not =0$.

Next, we compare the results with those obtained for the 
Schwinger model. In the abelian case, 
the combination of the matter boson $\varphi$ and the negative metric scalar 
$\eta$ gives rise to the $\theta$-angle. That is, the combination 
$\Phi\equiv\varphi+\eta=\theta $ 
appears in the mass term. When fermions are massless, the electric field
and the matter boson decouple. However, due to a Higgs mechanism, the 
electric field acquires a mass and, therefore, a long-range force does 
not exist. This leads to a pure screening potential. On the other hand,
for massive fermions, the electric field couples to the matter boson $\Phi$.
Yet, $\Phi=\psi_{c_V=0}$, and hence, it remains massless. This coupling
to $\Phi$ via the mass term is the origin of the long-range force (linearly
rising potential) in the massive U(1) case. Therefore the potential 
is confining. 

On the other hand, the expression (\ref{interquarkpotential}) for the 
potential indicates the absence of a long-range force in the non-abelian 
case. This can be understood by recalling that in such a case 
$\Phi\not=\psi$, so that $\Phi$ describes a massive field. The massless 
field $\psi$ decouples from the electric field (see eqn. 
(\ref{11massiveeqe})). The massive field
$\Phi\equiv\varphi+\eta$,  is the combination
that couples to $E$. Therefore, as both $E$ and $\Phi$ are massive, there 
is no long-range force. This is confirmed by our explicit computations.

The abelian potential can also be obtained from (\ref{interquarkpotential})
by taking the limit $c_V\to 0$. In this limit, the mass scale 
$m_-$ tends to zero and we recover the linearly rising potential, 
signaling confinement.

It is interesting to examine the behaviour of the screening
potential (\ref{interquarkpotential}) in extreme limits. 
In the strong coupling regime, $\lambda^2>>m^2$, the mass parameter 
$m_+$ dominates ($m_+>>m_-$) and we have 
\be
V(L)_{(m<< e)}\simeq {(c_V+1)^2 q^2\over 
2}\left\{ {(1-e^{-2\sqrt\pi\lambda
L})\over 2\sqrt\pi \lambda}+{\sqrt{\pi\epsilon } m\over
\lambda^2}(1-e^{-2\sqrt{2\pi\epsilon} mL}) \right\}.
\ee
On the other hand, in the weak coupling limit, $m>>e$, we obtain
\be
V(L)_{(m>>e)}\simeq{(c_V+1)q^2\over 
4\sqrt\pi\lambda}\sqrt{{1+\epsilon\over\epsilon}}
\left(1-e^{-2\sqrt{\pi\epsilon/(1+\epsilon)}\lambda L}\right)\quad .
\ee
In both regimes, the potential is governed by the parameter $\lambda$,
{\it i.e.} by the coupling constant.

\subsubsection{Adding flavour}
So far, we have considered fermions in the fundamental representation of 
$U(N)$. Explicit bosonisation formulas for fields in a higher representations
are in general not available. We can however proceed by introducing $F$ copies
$\{ \psi^i_f\}=\psi_1^i\psi^i_2,\cdots\psi_F^i$ of the $U(N)$ fermionic 
fields, labelled by a flavour quantum number $f$. We then treat the mass 
term perturbatively, introducing it via the Coleman's 
principle of form-invariance. The corresponding effective action is a 
simple generalization of (\ref{massive-boson-action})), with a set of 
$F$ matrix valued fields $g_f, f=1,\cdots F$, each in the fundamental 
representation of $SU(N)_L\times SU(N)_R$.

Following the procedure of the last section, we fix the external charges 
in colour space, parametrise the fields as in equation (36) and take the 
weak-field and static limit, to arrive at the same conclusions as before, 
namely, at the screening phase.
\subsubsection{Exotic states}
In the preceding sections, we have performed a semi-classical analysis
in order to understand the mechanism of screening and confinement in
two-dimensional QCD. In order to distinguish between the different phases,
we have used a dipole-dissociation test. If the
particles are confined, an infinite amount of energy is required to 
isolate them. In this case, as the inter-quark distance increases,
pair production occurs which obscures the physical interpretation of the 
results. On the other hand, in the screening phase the amount of 
energy required to dissociate the dipole is finite. Although charge (or 
colour) cannot be seen because of vacuum polarisation, further structures 
(or quantum numbers) can be observed.

In this section, we outline the construction of eigenstates of the 
hamiltonian which carry flavour quantum numbers. These are the analogues 
of the exotic states in the Schwinger model\cite{screening1}. Such a 
discussion provides a more elaborate confirmation of the screening phase.

We construct the exotic states by means of the fermionic operator
\cite{aar}.
\be
{\cal F}_f(x)=\prod_a 
e^{i\sqrt\pi\phi_f^a(x^0,x^1)/(c_v+k)+
i(c_v+k)\sqrt\pi\int_{-\infty}^{x^1}\dot\phi_f^a (x^0,y^1) 
dy^1} =\prod_a{\cal F}_f^a ,\label{exotic}
\ee
where the field $\phi_f^a$ does not carry colour charge.\footnote{We do 
not expect (\ref{exotic}) to be the complete operator which describes
flavoured physical states. Corrections involving multiple commutators,
due to the non-abelian character of the theory, can appear.} From the
semi-classical discussion, the combination $(\varphi^a_f+\eta)$ is the
natural candidate for the operator $\phi^a_f$. This is because we have
chosen symmetric boundary conditions $(\phi^a_f(+\infty)=\phi^a(-\infty))$
which imply that $\phi^a_f$ carries no charge. In the quantum 
theory\cite{ar-plb}, the operator $\phi^a_f=\varphi^a_f+\eta$ appears 
in the BRST current
\be
J_+=c_+\left(ig^{-1}_f\partial_+ 
g_f-i(c_v+k)\Sigma^{-1}\partial_+\Sigma+{\rm ghosts}\right),
\ee
which is conserved (actually vanishing) and leads to the topological charges
\be
Q=\left(\sum_{f=1}^k\varphi_f+(c_v+k)\eta\right)(t,\infty)-
\left(\sum_{f=1}^k\varphi_f+
(c_v+k)\eta\right)(t,-\infty)+\cdots\,.
\ee
where the dots stand for commutator-type corrections.

Next, we argue that the operator (\ref{exotic}) commutes with the mass term. 
We use the parametrisation
\begin{eqnarray}
g&=&e^{i\varphi^1\sigma^1}e^{i\varphi^2\sigma^2}e^{i\varphi^3\sigma^3},\\
\Sigma&=&e^{-i\eta^1\sigma^1}e^{-i\eta^2\sigma^2}e^{-i\eta^3\sigma^3},
\end{eqnarray}
in the SU(N) model and take the commutator of ${\cal F}_f$ with the mass 
term. This shifts $\varphi^a$ by 
$2\pi(c_v+k)$, and $\eta^a$ by $2\pi$. Since SU(2) is a compact group,
we conclude that ${\cal F}_f$ commutes with the hamiltonian.
This result can be generalized to any SU(N) gauge group.

By comparing expression (\ref{exotic}) with the fermionic operator,
we see that the field $\eta$ plays a r\^ole similar to that played 
previously by the sum $\sum_{i=1}^k \psi_k$ in the abelian theory. 
Consequently, the fields are not constrained in the non-abelian model 
and enjoy canonical commutation relations. Thus, kink dressing might 
be needed \cite{rothe1,rothe2}. In 
addition, the $\theta$-vacuum does not enter the expression for the 
fermionic operator (\ref{exotic}) in the non-abelian theory.

\subsubsection{Validity of the semi-classical approach and prospects about 
the four-dimensional theory}
The discussion of previous sections is based on a semi-classical approach
to QCD$_2$, which in view of the subtleties
linked to the quantum behaviour of the theory, may lead to doubts
about the validity of the method, especially concerning
application to the four dimensional case. However we shall argue that
we get the correct picture in the two-dimensional case
and that the problem can be pursued in four dimensions.

First some brief prolegomena, which concern the bosonisation procedure.
It is well known that the bosonised theory contains quantum information
at the classical level. Indeed, the anomaly equation, which is a one-loop
effect in the fermionic theory, is contained in the classical field
equation of the bosonic field. We also add the fact that most interesting
effects concerning two-dimensional gauge theories are one loop effects,
such as mass generation for the gauge field (coming directly from the
anomaly equation, thus a classical effect in the bosonic language)
or the vacuum structure, which in the bosonic language arises from
the rather intuitive Karabali-Schnitzer arguments\cite{ar-plb,kara-sch}.

However, the next indication is stronger, since it is based on real 
computations, and concerns the Schwinger model case. We generally
find a confinement potential, which disappears for a particular 
$\theta$-world and for massless fermions. Such a computation is done using
the same semi-classical reasonning advocated afterwards in the non-abelian 
case. The full quantum theory subsequently confirms this picture, by means 
of the construction of exotic states, which only commutes
with the mass term for the particular $\theta$-world where screening
prevails, confirming screening in that and in the massless cases. For a
detailed discussion and abundant literature concerning those points,
see chapter 10 of reference \cite{aar}. This settles the question
in full for the Schwinger model.

In the non-abelian case, although not solvable from first principles,
the situation is analogous. After finding the screening potential in
eq. (\ref{interquarkpotential}), with the various limits correctly 
reproduced, we have been
able to construct the state (\ref{exotic}), which commutes with the mass term,
being thus observable. This puts the non-abelian case in pair with the
Schwinger model, since the operator described realizes the screening
picture, albeit not being the most general quantum solution, which is however
beyond the scope of the discussion, since for our aim it is not necessary
to display the full set of exotic states, one single being enough to
confirm the issue. One further confirmation of the result consists in
noticing that the exotic state thus constructed is trivial 
in the abelian case, since the combination $\varphi +\eta$ acts as a constant
there. Thus, the screening phenomenon is strictly non-abelian.

The generalization of the method to the 3+1 dimensional case requires a
detailed knowledge of the bosonic form of the action, which as commented 
before, contains quantum information at the classical level. Therefore,
although technically much more complicated, the method itself
has chances of being applicable. Since we are dealing with static solutions
of the equations of motion, the results are presumably at least as
trustful as the instanton gas formulation of QCD.

One possible source of corrections is to consider the external charges as
obeying a certain dynamics, being classical solutions of field 
equations\cite{ellis}. The present methods as applied to that case
do not differ in any fundamental way from what we presented here.


\section{Integrability}
\subsection{Equations of motion and integrability condition}

We are now going to deal with the action $S_{eff}$ given in 
(\ref{ch11localwpf}), obtaining further important information.
Due to the presence of higher derivatives in that action, it is
convenient to introduce an auxiliary field and rewrite it in the equivalent 
form (\ref{ch11ymaction1}), or else (\ref{ch11ymaction2}).

The equation of motion of the $W$-field is easily computed. The WZW 
contribution has been obtained in \cite{ar-prd}, and the Yang-Mills action
leads to an extra term. We obtain
\bear
&&{c_V+1\over 4\pi}\partial_+\left( W\partial_-W^{-1}\right) +
{1\over (4\pi\mu)^2}\partial_+\partial_-\left( W\partial_-W^{-1}\right)
-{1\over (4\pi\mu)^2}\partial_+\lbrack W\partial_-W^{-1},
\partial_+\left( W\partial_-W^{-1}\right)\rbrack\equiv\nonumber\\
&&{c_V+1\over 4\pi}\partial_+\left( W\partial_-W^{-1}\right) 
+{1\over (4\pi\mu)^2}\partial_+{\cal D}_-\left( W\partial_-W^{-1}\right)=0
\label{eqmotionw}
\eear

We can  list the relevant field operators appearing in the definition of the
conservation law (\ref{eqmotionw}), that is
\be
I^W_-= {1\over 4\pi} (c_V +1) J^W_- + {1\over (4\pi\mu)^2} \partial_+ 
\partial_-J^W_- - {1\over (4\pi\mu)^2} [ J^W_-, \partial_+J^W_-]\quad ,
\label{conslaww}
\ee
with $\partial _+  I^W_- = 0$, and $J^W_- = W\partial _-W^{-1}$.
It is straightforward to compute the Poisson algebra, using the canonical
formalism, which in the bosonic formulation includes quantum corrections. 
We have
\bear
\left\{I^W_{ij}(t,x), I^W_{kl}(t,y) \right\}  &=& \left[ I^W_{kj}
\delta_{il} - I^W_{il}\delta_{kj}\right]\delta(x^1\!-\!y^1)-\alpha\delta^{il}
\delta^{kj}\delta'(x^1-y^1)\quad,\nonumber\\
\left\{ I^W_{ij}(t,x), J^W_{-kl}(t,y) \right\}&=& (J^W_{-kj}\delta_{il} - 
J^W_{-il}\delta_{kj})\delta(x^1-y^1)
+ 2\delta_{il}\delta_{kj}\delta'(x^1-y^1)\quad ,\label{fullpoissonconslaw}\\ 
\left\{ J^W_{ij}(t,x), J^W_{-kl}(t,y^1) \right\}&=& 0\quad .\nonumber
\eear
where $\alpha = {1\over 2\pi}(c_V+1)$.
We thus obtain a current algebra for $I_-^W$, acting on $J^W_-$ with a central 
extension.

\subsection{Dual case-non local formulation}
At the Lagrangian level, we find the Euler-Lagrange equations for
$\beta$ from the perturbed WZW action (\ref{zbeta}), that is,
\bear
\delta \Gamma[\beta ]&=& \left[{1\over 4\pi}\partial_-(\beta^{-1}\partial_+ 
\beta)\right]\beta^{-1}\delta\beta\quad ,\\ 
\delta\Delta(\beta) & =& 2\Big( \partial_+^{-1} 
(\beta^{-1}\partial_+\beta)  - \big[ \partial_+^{-2} 
(\beta^{-1}\partial_+\beta),(\beta^{-1}\partial_+\beta)
\big] \Big)\beta^{-1}\delta \beta\quad .
\eear

We define the current components
\bear
J_+^\beta&=& \beta^{-1}\partial_+ \beta\quad ,\\ 
J_-^\beta&= &-4\pi\mu^2\partial_+^{-2}J_+^\beta=-4\pi\mu^{2} 
\partial_+^{-2}(\beta^{-1}\partial_+\beta)\quad ,\label{currentbeta}
\eear
which summarize the $\beta $ equation of motion as a zero-curvature condition
given by
\be 
[{\cal L},{\cal L}]=[\partial_++J_+^\beta,\partial_- +J_-^\beta]=\partial_- 
J_+^\beta-\partial_+J_-^\beta+[J_-^\beta, J_+^\beta ] =0 \quad .
\label{laxpaircandidate}
\ee 
This is not a Lax pair, as {\it e.g.} in the usual non-linear $\sigma$-models, 
where $J^\beta_\mu$ is a conserved current and a conserved 
non-local charge is obtained. However, to a certain extent, the 
situation is simpler in the present case, due to the rather unusual form of 
the currents, which permits us to write the commutator
as a total derivative, in such a way that in terms of the current 
$J_-^\beta$ we have
\be 
\partial_+\left(4\pi\mu^2J_-^\beta+\partial_+\partial_-J_-^\beta+[J_-^\beta,
\partial_+J_-^\beta]\right)=0\quad .\label{consvlawbeta}
\ee

Therefore the quantity
\be 
I_-^\beta (x^-) = 4\pi \mu^2 J_-^\beta (x^+,x^-) + \partial_+ \partial_-
J_-^\beta (x^+,x^-) + [J_-^\beta (x^+,x^-), \partial_+ J_-^\beta(x^+,x^-)]
 \quad \label{iconservedbeta}
\ee
does not depend on $x^+$, and it is a simple matter to derive an infinite 
number of conservation laws from the above.

Canonical quantization proceeds straightforwardly, and as a consequence we can
compute the algebra of conserved currents, which is analogous to
(\ref{fullpoissonconslaw}).

This means that two-dimensional QCD contains an integrable 
system\cite{aar,faddeev-gn}. Moreover, it corresponds to an 
off-critical perturbation of the WZW action. If we write $\beta=
\E^{i\phi}\sim 1+i\phi$,  we verify that the perturbing term 
corresponds to a mass term for $\phi$. The next natural step is to obtain 
the algebra obeyed by (\ref{iconservedbeta}), and its representation. 
However, there is a difficulty presented by the non-locality of the 
perturbation.

\section{ Algebraic aspects of QCD$_2$ and integrability}
We saw that two-dimensional QCD, although not exactly soluble, in 
terms of free fields, is a theory from which some valuable results may be 
obtained. 
The $1/N$ expansion reveals a simple spectrum valid for weak coupling, while 
the strong coupling offers the possibility of understanding the baryon as a 
generalized sine-Gordon soliton. Moreover, the $1/N$ expansion of the 
pure-gauge case may be performed, and the partition function is equivalent to  
one of a string model described by a topological field theory, the Nambu-Goto 
string action, and presumably terms preventing folds.                       

All such results point to a relatively simple structure, which could be
mirrored by an underlying symmetry algebra. In fact such algebraic structures 
do exist. In the above-mentioned case of the large-$N$  expansion of pure 
QCD$_2$, one finds a $W_\infty$-structure related to area-preserving 
diffeomorphisms of the Nambu-Goto action. A $W_\infty$ structure for 
gauge-invariant bilinears in the Fermi fields can be constructed \cite{dmw}. 
Such is an algebra which appears also in fermionic systems, and 
in the description of the quantum Hall effect. Moreover, 
pure QCD$_2$ is equivalent to the $c=1$ matrix model,  which 
also has a representation in terms of non-relativistic fermions,  and 
contains a $W_\infty$ algebra  as well. The problem is also 
related to the Calogero-Sutherland models.  The mass eigenstates build
a representation of the $W_\infty$ algebra as found in \cite{dmw}.

After bosonizing the theory, further algebraic functions of the fields turn out
to obey non-trivial conservation laws, as we will see. The theory can be 
related to a product of several conformally invariant WZW sectors, a perturbed 
WZW sector, all related by means of BRST constraints, which play a very 
important role in gauge theories. A dual formulation exists and permits us
to study the theory in both strong and weak coupling limits. Finally,
once displayed, the relation to Calogero systems and further integrable 
models is also amenable to understanding in the previous framework.

\section{Conclusions}
In twenty years of development, two-dimensional QCD made outstanding 
contributions to the non-perturbative comprehension of strong
interactions. The large-$N$  limit of the theory revealed a desirable structure
for the mesonic spectrum, whose higher levels display a Regge behaviour. These 
properties were generalized for fermions in the adjoint representation. This
is an important step towards understanding the theory in higher dimensions,
since adjoint matter can substitute the lack of the transverse
degrees of freedom of the gauge field in two dimensions. 
Matter in the adjoint representation of the gauge group provides fields which 
mimic the transverse degrees of freedom characteristic of gauge theories in 
higher dimensions, and may show more realistic aspects of strong interactions. 
The main new point consists in the presence of a phase transition indicating a 
deconfining temperature. Further properties of
the perturbative theory are also in accordance with expectations for strong
interactions, and therefore it has certain advantages over the
usual non-linear $\sigma$-models in the description of strong interactions by
means of simplified models. The large-$N$ limit of QCD$_2$ is smooth and provides a picture of the string in the Feynman diagram space.  In certain 
low-dimensional systems, the $1/N$ expansion turns out to be the correct
expansion. In models with problematic infrared behaviour, such as $\Complex 
P^{N-1}$ and Gross-Neveu models, properties such as 
confinement and spontaneous mass generation are easily obtained in 
the large-$N$ approximation and the $S$-matrices are explicitly 
checked\cite{cpn-gn}.

The computation of the non-Abelian fermionic determinant is the key 
to tacking the problem of confinement which provides an effective theory 
for the description of the mesonic bound states, and to opening the 
possibility of understanding baryons as solitons of the effective 
interactions. Major problems of QCD can be dealt with using these methods. 

The string interpretation of pure Yang-Mills theory, as well as its
Landau-Ginzburg-type generalizations connects the previously mentioned 
picture to that of non-critical string theory. These developments form
the basis for a deeper understanding of the r\^ole of non-critical string
theory in the realm of strong interactions. String theory although far from 
being realized, seems to be the correct way to understand strong interactions at
intermediate energies. For full details see \cite{kut-schw}.

The question of high-energy scattering in strong interactions is
linked with two-dimensional integrable models. Thus the higher-symmetry
algebras, spectrum-generating algebras, and integrability conditions, might
give clues to the understanding of two-dimensional QCD. At high energies,
Feynman diagrams simplify
and become effectively two dimensional. The theory may be described in the
impact parameter space, and in the case of QCD$_4$, the Reggeized particles
scatter according to an integrable Hamiltonian.

In the massless case, the external field problem for the effective action
in particular can be 
analyzed and  the computation of gauge current and fermionic Green's 
functions can be reduced to the calculation of tree diagrams. These
features are also 
not covered by 't Hooft's method. As an example, if we take the 
pseudo-divergence of the Maxwell equation, i.e.,
\be 
\left({\cal D}^2 + (c_V+1) {e^2\over \pi}\right) F = 0 \quad ,    
\ee
where $F=\epsilon_{\mu\nu}F^{\mu\nu}$. This equation generalizes the analogous 
result obtained for the Schwinger model to the non-Abelian case. This suggests 
that an intrinsic Higgs mechanism, analogous to the one well-known in QED$_2$, 
can also characterize the non-Abelian theory. This is, nevertheless, not
contained in 't Hooft's approach, since the mass arises from the fermion loop,
which also contributes to the axial anomaly. This is suppressed in the 
$1/N$ expansion.

In spite of difficulties, QCD$_2$ has also served as a laboratory for gaining
insight
into various phenomenological aspects of four-dimensional strong interactions,
such as the Brodsky-Farrar scaling law for hadronic form factors, the
Drell-Yan-West relation or the Bloom-Gilman duality 
for deep inelastic lepton scattering.\cite{scaling-test} 

The next important step towards understanding this theory is its relation to
string theory, or SCD$_2$. It concerns one of the most important applications 
of the theory of non-critical strings.\cite{aadz}

The general problem of strong interactions did not progress substantially until
recently as far as it concerns low-energy phenomena. Such a problem should be
addressed using non-perturbative methods, since perturbation theory of strong
interactions is only appropriate for the high-energy domain, missing
confinement, bound-state structure and related phenomena. In fact, several
properties concerning hadrons are understandable by means of the concept of
string-like flux tubes, which are consistent with linear confinement and Regge
trajectories, as well as the approximate duality of hadronic scattering
amplitudes, which are the usual concepts of the string idea. In fact, a
similar idea is already present in the construction of the dipole of the
Schwinger model, in which case it is, however, far too simple to be realistic.

In short, these ideas support the suggestion  that the understanding of the 
theory of strong interactions requires the study of the large-$N$ limit of QCD.
Although there are several candidate models for such a theory in two 
dimensions, concerning a thorough comprehension of subtle problems such as
confinement cannot be obtained without the inclusion of QCD$_2$.
\chapter{Two-dimensional gravity}

\section{Introduction }
String theory is intensively discussed in modern quantum field theory.
It has been under focus for the past thirty  years, had a 
complex development, served different purposes and has undergone
major changes.

The theory emerged from  the dual models whose aim was to present an 
alternative to quantum field theory, one which would not rely on 
perturbative schemes. In particular, the failure of perturbative approach 
to strong interactions, was the main motivation of dual models and
subsequently the initial objective of string theory.
Due to the lack of a dynamical principle, the dual models
were not predictive enough to be 
tested experimentally. Nevertheless many ideas fructified, as e.g. the concept 
of duality and the Veneziano formula, which later permitted a reconciliation 
of the dual models with quantum field theory by means of the introduction of 
the concept of string dynamics.

With the advent of supersymmetry and the GSO projection, string 
theory, which initially aimed at the explanation of strong interactions was 
then targeting a more ambitious task: the formulation of a 
unified theory of all interactions.

Thus, one sees already two aspects of the string models, the first
concerns strong interactions where non-critical strings is used to 
describe the physical space-time which, in principle, cannot 
be studied by the usual string dynamics. In physical dimensions,
string theory is anomalous and a Wess Zumino term has to be 
introduced to restore the symmetries. Non-critical strings incorporating
the Wess-Zumino term are valuable for studying various 
statistical mechanical problems and have lead to the development of 
matrix models, which have applications in areas such as nuclear physics, 
and transport phenomena.

The second aspect of strings, that which presents it as a model of all 
interactions, still
follows a very refined and mathematically sophisticated path. Today,
theories of strings and superstrings are believed to contain higher 
symmetries which relate different types of strings. Such is the duality 
symmetry, which also lead to the re-newed interest on extended 
objects: the membrane (or 
more generally the p-brane) theories. Here, we shall concentrate on the 
problems related to two-dimensional gravity as a model, and partially 
as a tool for the understanding of non-critical string theory, without
getting into the specific details of string theory.
\section{Polyakov formulation of the Nambu-Goto string action}
The classical dynamics of strings is entirely based on geometry. The 
Nambu-Goto action is proportional to the area swept by the string as it 
evolves in space-time, that is 
\be
S_{NG}= \lambda \int {\rm  d} ^2 \xi \sqrt{(\dot X \cdot X')^2 - \dot 
X^2 {X'}^2}= \lambda \int {\rm d}  ^2 \xi 
(\det \gamma_{ab})^{1/2}\quad , \label{nambuaction}
\ee
where $X^\mu(\xi_0,\xi_1)$ describes the position of the string and
\be
\gamma_{ab}= \partial _a X^\mu \partial _bX_\mu\label{17inducedmetricgamma}
\ee
is the induced metric.

This action describes a  field $X^\mu$ which obeys the minimum area equation
$\partial _a[(\det \gamma)^{1/2} \gamma^{ab} \partial
_b x^\mu]=0$. This equation is equivalent to the
two-dimensional Klein-Gordon equation, supplemented by the so-called 
Virasoro conditions which is in turn obtained by requiring the metric to
be diagonal. It is not difficult to see that this same set of conditions is
obtained from a free field action for $X^\mu$ which incorporates a
two-dimensional gravitational field $g_{\alpha \beta }$. Such
is the Polyakov\cite{polyakov-spr} string action
\be 
S = {1\over 2} \int {\rm d} ^2 \xi \sqrt {|g|} g^{ab}\partial _a 
X^\mu \partial _b X_\mu\quad ,
\label{polyakovaction}
\ee
where $\xi=(\xi^1,\xi^2)$ are local coordinates on $M, X^\mu (\xi),
(\mu=1,...,D)$  defines the embedding from the world-sheet to the
$D$-dimensional space-time, and  $g  = \det g_{ab}$. For reasons to
become clear later on, $D$ is taken to be arbitrary for the time being.

In the Polyakov formulation, the connection of string theory with 
two-dimensional gravity becomes most transparent. Indeed, $M$ is a
compact orientable two-dimensional manifold with boundary $\partial M$, and
with metric tensor  $g_{ab}$. In such a case,  
$X^\mu(\xi):M\to {\rm l}\!{\rm R}^D$
is an embedding of $M$ in $D$-dimensional Minkowski  space.
The classical Polyakov action is then given by (\ref{polyakovaction}).

\section{The effective action of quantum gravity}
We now proceed to discuss the quantization of the theory defined by the
Polyakov action (\ref{polyakovaction}). The classical action is invariant
under reparametrizations as well as under Weyl transformations. However
at the quantum level only one of these symmetries can be maintained.
This is analogous to the situation which arises in gauge theories, 
where either vector or axial vector
current conservation can be implemented at the quantum level. We shall 
maintain reparametrization invariance, at the expense of loosing Weyl 
invariance. Renormalization will in general require the introduction of
counter terms, which  in general complicate the structure of a simple
diffeomorphism invariant action. Making use of this invariance,
we proceed to quantize the theory in the conformal gauge. Taking into
account the Faddeev-Popov determinant and  after
integrating over the fields $X^\mu$ of the ``target" space, we arrive at an
effective action of the ``Liouville" type. This action has a coefficient 
which only vanishes in $26$ dimensions. This shows 
that in the critical dimension Weyl invariance can also be
implemented and consequently the dynamics is that of a free field.
In $D\not =26$, the breakdown of Weyl invariance leads to a non-trivial
dynamics.
\subsection{Uniqueness of the Polyakov action}
At the classical level, the Polyakov action (\ref{polyakovaction}) is 
not the most general action compatible with all symmetries of the theory. The 
requirement of Poincar\'e invariance implies that the corresponding 
Lagrangian can only depend on the derivatives of $X^\mu$, since the 
translations $X^\mu\to X^\mu+a^\mu$  must represent a symmetry of the action; 
the further requirement of diffeomorphism invariance permits the following
generalized form of the action
\be
S'[x,g]{1\over2}A\int_Md^2\xi\sqrt{-g}g^{ab}\partial_a X^\mu\partial
_b X_\mu
+{1\over{2\pi}}B\int_Md^2\xi\sqrt{-g}R+\mu^2\int d^2\xi\sqrt{-g}
\label{genpolyakov}
\ee
where $A$, $B$ and $\mu^2$ play the role of renormalization constants,
and $R$ is the Ricci curvature scalar. We refer to (\ref{genpolyakov})
as the Polyakov-Zheltukin action.

The second and third terms in the action (\ref{genpolyakov}) are the
Einstein action (without matter fields) and the cosmological term,
respectively. The former does  not contribute to the equations of
motion, but merely modifies the boundary conditions (which we do not
consider in this sequel). The equation of motion associated
with the Einstein action
\be
S_{Einstein}=\int d^2\xi\sqrt{-g}R\label{einstein}
\ee
reads
\be
R_{ab}-{1\over2} g_{ab}R=0\label{einsteineq}
\ee
and is trivially satisfied, since the two-dimensional geometry leads to
the identity $R_{ab}\equiv {1\over2} g_{ab} R$. The Gauss-Bonnet
theorem states that Einstein's Lagrangian is a total derivative in two 
dimensions. This means that the two-dimensional action is a topological
invariant quantity which
measures the  genus  of the manifold on which one integrates. 
For a  compact manifold  $M$ with $b$  boundaries one has
\be
S_{Einstein}=\int d^2\xi\sqrt{-g}R=2\pi n=2\pi (1-g)\label{gaussbonnet}
\ee
where $g$ is the genus of the manifold and the number $2(1-g)$  is 
its Euler characteristic. Since it is a topological invariant
quantity, the Euler characteristic does not participate in the equation
of motion, which turns out to be a
constraint for the gravitational field ($T_{ab}$ (matter) = 0), rather than
a dynamical equation. This is the case in string and superstring theories
at the critical dimension (where also Weyl invariance holds). Indeed, the
graviton (and gravitino) equations of motion ensure reparametrization
(and super-reparametrization) invariance by means of the Virasoro
(and super Virasoro) conditions.
\subsection{Quantum Gravity}
We shall now proceed with the quantization of the Polyakov model.
Following Polyakov, we shall do this within the functional framework.
Restricting ourselves for the time being to manifolds without boundaries
and handles, we consider the expression
\be
{\cal Z}=\int {\cal D} g\int {\cal D} X^\mu 
\E ^{-\int d^2\xi\sqrt g g^{ab}\partial_a X^\mu\partial_b
X_\mu}\quad .\label{partgrav}
\ee
For the Euclidean  partition  function associated with the Polyakov action 
$A=2$, $B=\mu^2 =0$). This  is a merely formal expression, since as a result 
of the reparametrization invariance of the action,  this functional integral
is actually infinite. In order to define a finite integral we shall
have to adopt the Faddeev-Popov procedure. At the classical
level, Polyakov's action is proportional to the area of a
two-dimensional surface on the manifold $M$. We are thus faced with the
problem of summing over surfaces randomly immersed in
a $D$-dimensional Euclidean space. The analogous problem of one-dimensional
curves randomly immersed in the Euclidean space corresponds to a system of
Brownian motion and to the quantum theory of a free relativistic point
particle after continuation to Minkowski space. It is thus natural that 
in the case in question we are lead to the quantum theory of strings.

To begin with we need to define the measure in (\ref{partgrav}).
Polyakov has discussed this question in detail.

\subsubsection{Going to the conformal gauge}
Since the action (\ref{genpolyakov}) is diffeomorphism invariant Faddeev-Popov
procedure is required for the evaluation of the partition function.

The first step in this procedure is taken by choosing a gauge. 
In particular, on a sphere (no handles and no boundaries) we may
always choose the conformal gauge  in which the metric tensor takes the form
\be
g_{ab}(\xi)=e^{\phi(\xi)}\eta_{ab}\quad .\label{confgauge}
\ee

This gauge can only be chosen if we perform a suitable 
conformal transformation. There are however topological obstructions.

\subsubsection{Computation of The Faddeev-Popov determinant}
We can complete the derivation of the effective action of quantum gravity
by explicitly computing the Faddeev-Popov determinants
appearing in the measure (\ref{partgrav}). The
corresponding heat kernel is expected to have a de Witt-Seeley expansion.
Combining the results of both matter and ghost calculations, 
we arrive at the final form of the effective action, which is given by
\be
S^{eff}_{tot}={{26-D}\over{48\pi}}\int d^2\xi\left[\delta^{ab}\partial_a\phi
\partial_b\phi+\mu^2e^{-\phi}\right]\label{liouvd26}
\ee
We shall refer to $S_{eff}$ as the Liouville action.

\section{The Liouville theory}
The classical Polyakov action is invariant under reparametri\-zation and 
Weyl transformations.
Quantization breaks Weyl-invariance, but preserves conformal
invariance. This allows one to choose the conformal gauge in order to
simplify the calculations. After gauge fixing on the orbits defined
by the reparametrization, the infinite group volume corresponding to
the conformal transformations can be factored out at the expense of a
Faddeed-Popov determinant represented in terms of a ghost action of
the Liouville type. On the other hand, the one-loop effective action
breaks the Weyl invariance, and hence shows a non-trivial dependence
on the conformal factor in the metric. This dependence is once again, of the
Liouville type. In the critical dimension $D = 26$, the one-loop contributions
of the original action and the ghost action cancel out. Hence, a
further gauge fixing is required, by means of which the group volume
corresponding to the Weyl transformations can once more be factored out. 

The situation witnessed here is in fact analogous to that
occurring in anomalous chiral gauge theories in the 
gauge-invariant formulation. The chiral invariance of the effective
one-loop action is restored in that case by the Wess-Zumino term.
In the present case this r\^ole is taken up by the Liouville action
associated with the Faddeev-Popov term of diffeomorphism invariance.

In the absence of the Weyl anomaly the partition function (\ref{partgrav})
of quantum gravity describes a purely topological theory which is 
characterized by the genus of the manifold. It is the 
breakdown of Weyl invariance in the non-critical dimension ($D\not =26$) which
adds a term of the form (\ref{liouvd26}) to the action, thereby
rendering the dynamics non-trivial.

\subsection{The classical Liouville theory}
The Liouville equation has been known for more than a century. In classical 
mathematics it has been used in order to study the uniformization problem by
Poincar\'e. A knowledge of Liouville theory is essential for understanding
this important mathematical problem.

The generalized Liouville action is
\be 
S={1\over 4\pi}\int {\rm d}^2 z \sqrt {|\hat g|}\left\{ {1\over 2}\partial
^\mu \phi \partial _\mu \phi +{2\mu \over \gamma ^2}\E^{\gamma\phi }+
{1\over \gamma }\phi \hat R(\hat g)\right\} \quad ,\label{liouncrit}
\ee 
where $\hat g$ is a fiducial metric and we have found it convenient to rescale
the Liouville field as $\phi\to\gamma\phi$, with $\gamma =\sqrt{
c\over 6\pi}$. The last term in (\ref{liouncrit}) vanishes in flat space.

In the flat space ($\hat R=0$), the equation of motion of the Liouville field 
has the general solution
\be 
\E^{\gamma\phi(z,\overline z) }={4\over \mu }{A'(z)B'(\overline z)\over 
(1-A(z)B(\overline z))^2}\quad ,
\label{solliouncrit}
\ee
where the prime indicates differentiation with respect to the arguments.

On the other hand, on the space of a punctured Riemann sphere it is always 
possible to find a conformal map such that the curvature is localized at 
isolated singular points. In this case, the curvature $\hat R (\hat g)$
is given in terms of delta functions. Thus (\ref{solliouncrit}) continues
to be a solution away from the singularities.

Let us consider the case of one such a singularity at the origin of
the $z$-plane. Neglecting for the moment the cosmological term, the Liouville 
field obeys the equation,
\be 
\Delta \phi +{1-a\over \gamma}8\pi
\delta^{(2)}(z)=0 \quad ,\label{freelioucurv}
\ee
with the solution
\be 
\phi = {a-1\over \gamma }\ln (z\overline z)\quad , \label{philnzzbar}
\ee 
where $a$ parametrizes the strength of the curvature at the origin. It is 
essential to check that neglecting the cosmological term does not affect the
above solution near the singularity. From (\ref{philnzzbar}) we have
\be 
\E^{\gamma\phi } \sim {\vert z\vert ^{-2(1-a)}}\quad ,\label{singphi} 
\ee 
which is integrable only for $a>0$. This forbids the existence of
highly-curved single points. As a matter of fact, the results below show
that the classical solution to the field equations exactly matches the above
condition. This result will also be used to characterize the
conditions to be fulfilled by the constants in the quantum theory.

Next we include the cosmological term and study the different solutions of 
the generalized Liouville equation (we only include one source of curvature)
\be 
\Delta \phi -{2\mu \over \gamma }\E^{\gamma \phi }+{1-a\over \gamma}8\pi
\delta^{(2)}(z)=0 \quad ,\label{lioucurv}
\ee
and obtain solutions of the type (\ref{solliouncrit}) with $A(z)=z^a$ and
$B(\overline z)=\overline z^a$,
\be
\E^{\gamma \phi }={4\over \mu }{a^2\over (z\bar z)^{1-a}[1-(z\bar
z)^a]^2}\quad .\label{asympsol}
\ee

This concludes our summary of classical Liouville theory.

\subsection{The quantum Liouville theory}
Quantum Liouville theory has been studied intensively (see \cite{aadz}).
It has been discussed as a classical integrable model with boundary conditions
\cite{ger-neveu1}, and the string spectrum has been analysed\cite{ger-neveu2}.
The full quantum operator solution has been studied in ref. \cite{bcgt} where  
conformal invariance has been considered in detail. The model including 
boundary terms was also studied in detail\cite{dur-n-o-p}.
  
The previous prolegomena warrant the importance of the Liouville theory 
in the study of two-dimensional random surfaces, in particular of 
quantum  gravity. As we shall see, it also serves as a useful device for
obtaining the exact correlation functions of the dressed vertex operator 
in non-critical string theory using the well known Coulomb gas 
method\cite{dot-fat}. Moreover, one is able to partially treat\cite{31b}
the Wheeler-de Witt 
equation\cite{wdw}, allowing for a better understanding of time in this 
simplified model, an ill defined concept in general relativity\cite{teitel}. 
Furthermore it is possible to compare different approaches to 
non-critical string theory, namely Liouville continuous approach at 
one hand, and matrix models (discrete) approach on the other hand. 

By integrating the equation of motion (\ref{lioucurv}) over the whole space, 
using $R=-\gamma\E^{-\gamma\phi}\Delta\phi$,
the Gauss-Bonnet theorem, and replacing ${1-a\over\gamma}$ by various external
sources of curvature of strength $\beta_i$, we obtain
\be
\gamma\sum\beta_i+2h-2-{\mu\over 8\pi}A=0\quad ,\label{intgaussbonnet}
\ee
where $h$ is the number of handles of the Riemann surface, and $A$ is its area.
This relation ensures that a classical solution exists if and only if
\be \gamma\sum\beta_i+2h-2>0\quad .\label{condgb}
\ee

To quantize the Liouville theory, we can use the canonical method. The energy
momentum tensor is traceless, i.e. $T_{+-}=0$, which implies conformal
invariance. We define the physical Hilbert space of states
by requiring the other components of the energy momentum tensor to be 
weakly zero. (The conformal dimension of the primary
operator $\E^{\beta\varphi}$ can be fixed.) 

The equation of motion obeyed by the exponential field can be shown to
be equivalent to the definition of a null vector in 
conformal field theory; it can be written as\cite{aadz}
\be 
\left(\partial _z^2 + \gamma^2 T(z) \right) \E^{-{1\over 
2}\gamma\phi}=0\quad , \label{phiisprimary}
\ee 
showing that $\E^{-{1\over 2}\gamma\phi}$ is a solution of a null vector
equation, thereby making contact with the methods of conformal field theory.

In order to have a better understanding of Liouville theory, we shall 
consider the Schr\"odinger problem associated with the Hamiltonian of the 
Liouville action, so that we can obtain (\ref{phiisprimary}) from first 
principles.
  
Let us  consider  the Lagrangian
\be 
{\cal L} = {\sqrt {\vert {\hat g}\vert}\over 4\pi}\left({1\over 2}
\hat g^{\mu \nu}
\partial _\mu \phi\partial _\nu \phi + {2\mu \over \gamma^2}\E^{\gamma \phi}
- {Q\over 2}\phi \hat R\right) \quad ,\label{lioubackgrav}
\ee 
where $Q$ is classically related to the coupling constant $\gamma$ by   
$Q=-{2\over \gamma}$. This relation will however acquire quantum corrections. 
The $(++)$ component of the energy momentum tensor defines a Hamiltonian as
\be 
H = {1\over 2} (\phi' + 4\pi P )^2 + Q(\phi' + 4\pi P )'
+ {\mu \over 2\gamma^2} \E^{\gamma \phi } + {Q^2\over 16}\hat R + 
{Q^2\over 8}\quad ,\label{hliouemtensor}
\ee 
where $P$ is the momentum canonically conjugated to $\phi$.
Note that this Hamiltonian is the generator of conformal transformations.
For flat space $\hat R=0$. In the so-called minisuperspace approximation,
where the quantities do not depend on the space variable $x^1$, the 
corresponding eigenvalue problem in the Schr\"odinger representation reads
\be 
H\psi\equiv\left(- {1\over 2} {\partial^2\over\partial\phi^2} + 
{\mu \over 2\gamma^2} \E^{\gamma \phi} + {Q^2 \over 8}\right) \psi 
=\Delta \psi \quad ,\label{schliou}
\ee
where $\Delta$ is the conformal dimension of $\psi$. Since the potential
vanishes exponentially for $\phi\to -\infty$, we have for the eigenfunctions 
of $H$
\be 
\psi\left( \phi\to -\infty \right) \approx \sin p\phi  \label{potinfty}
\ee 
where $p$ is the eigenvalue of $-i{\partial\over \partial\phi}$, and
$\Delta = {1\over 2}p^2 + {1\over 8}Q^2$.
However, there also exist eigen solutions corresponding to vertex operators 
of the form ${\cal O} \sim \E^{\beta \phi}$. These are not normalizable,
but are important in the context of non-critical string theory. 
Solving the Schr\"odinger Eq. (\ref{schliou}) we see that they are given by 
\be 
\psi_{{\cal O}}(\phi) = \E^{({1\over 2}Q+\beta)\phi}\label{solschliou}
\ee 
with conformal dimension
\be
\Delta=-{1\over
2}(\beta + {Q\over 2})^2 + {Q^2\over 8}\label{eigendeltaliou}
\ee
This is the general case in Liouville field theory as discussed by
Seiberg\cite{seiberg}.

Let us now consider the ``puncture" operator, obtained as a solution of Eq. 
(\ref{schliou}) with $\Delta = Q^2/8$. There are two solutions: $\psi =1$ 
and $\psi =\phi$, corresponding to the operators $ \E^{{Q\over 2}\phi}$ 
and $\phi\E^{{Q\over 2}\phi}$ respectively. 

An equation of motion of the form (\ref{lioucurv}) may be obtained from 
the study of correlation functions of the type
\be \left\langle\prod_i\E^{\beta_i\phi(z_i)}\right\rangle=\int {\cal D}\phi
\E^{-S[\phi ]}\prod_i\E^{\beta_i\phi (z_i)}\quad .\label{aveexpliou}
\ee
Therefore, in the quantum theory it is important to analyze the action of the 
operators $\E^{\beta\phi (z)}$ on the vacuum; these operators are responsible
for the local curvature strength $\beta$. That is to say that they generate
elliptic solutions such as (\ref{asympsol}), with $a=1-\gamma\beta$. 
From the discussion following (\ref{asympsol}) we see that we need
$\beta\gamma<1$, or \footnote{ $\gamma<0$, this is the case in the quantum 
theory, according to our conventions} $\beta > 1/\gamma$. 
Classically this is equivalent to requiring the inequality
\be 
\beta \ge -{Q\over 2}\quad ,\label{condonbeta}
\ee 
which continues to be correct even after taking into account quantum
corrections. Canonical quantization of the theory has been performed 
in \cite{br-c-th}, and one finds for the 
anomalous dimension of the above vertex operators
\be 
\Delta (\E^{\beta\phi})=-{1\over 2}\beta (\beta + Q)=-{1\over 2}
(\beta + {Q\over 2})^2+{c-1\over 24},\label{anodimexp}
\ee
where $c=1+3Q^2$ is the central charge of the Virasoro operator. Note that
(\ref{anodimexp}) is in agreement with (\ref{eigendeltaliou}) obtained in the
minisuperspace approximation. The result (\ref{anodimexp}) reflects the 
fact that Liouville theory cannot be
treated as a free theory. In fact, Liouville theory cannot be treated in
perturbation theory due to the lack of a normalizable ground state at finite
values of the Liouville field \footnote{for $\gamma\phi\to-\infty$ all 
derivatives drop to zero,  and the theory is trivial\cite{dho-jack}}. 

A comparison between the full action of string theory and the
Liouville field, shows that the latter  can be thought of as a target
space coordinate, and the full action is that of a string theory in a
non-trivial background. Thus, in quantum theory we shall sum over the possible
geometries. We consider the Liouville interaction with curvature and   
recall that in string theory the sum over geometries corresponds to a
string perturbation theory in terms of the genus, that is
\be 
Z \sim \sum Z_{h} g_{st}^{2\chi} \quad .\label{parstring}
\ee 
In addition, we use eq. (\ref{lioubackgrav}) for constant Liouville 
configurations and take into account the contribution from the fiducial metric
to the Euler characteristic of the manifold. Collecting these results together 
we find that the string coupling constant is related to the Liouville field by
\be 
g_{st} = g_0 \E^{-{Q\over 2}\phi}\quad .\label{gstring}
\ee 

The importance of this result will be shown later when we define the 
relation between the tachyon vertex
and the corresponding wave function,  for example in the discussion
concerning eq. (\ref{solschliou}).
\subsection{Canonical quantization and $SL(2,R)$ symmetry}
Liouville theory corresponds to two-dimensional gravity in the
conformal gauge. In order to obtain the corresponding formulation in the 
light-cone-gauge, we notice that the reparametrization invariant action
($\Box$ is the Laplace-Beltrami operator)
\be 
S_g = {c\over 96\pi}\int {\rm d}^2 x \lbrack \sqrt {\vert g\vert}R{1\over
\dal }R +\mu^2 \sqrt {\vert g\vert}\rbrack \quad ,\label{2dgrav}
\ee 
reduces to the Liouville action in the conformal gauge.

In order to study the symmetries of (\ref{2dgrav}) it is convenient to use
a formulation in terms of a local Lagrangian. We thus introduce an 
auxiliary field $\varphi $\cite{aar}
\be \varphi =-{\alpha \over 2}\dal ^{-1}R\quad ,\label{aux2dgrav}\ee
where $\alpha ^2=8\kappa $. Thus classically we have
\be \dal \varphi +{1\over 2}\alpha R=0\label{classphiaux}
\ee
and 
\be \kappa R{1\over \dal }R=-{1\over 2}(\varphi \dal \varphi
+\alpha R\varphi )\quad .\label{id2dgravaction}
\ee
We obtain correlators and operator-product expansions for the
elementary fields of the theory. The supersymmetric extension
is also amenable to the method, and analogous expressions have been
obtained in the literature. See \cite{aar} for an extensive literature.

\chapter{Four-dimensional analogies and consequences}

It is of prime importance to generalize the concept of 
quasi-integrability to higher dimensions. Indeed, Bardeen 
\cite{bardeen} has recently pointed out that helicity amplitudes in 
high-energy QCD
are very simple at tree level and are described by a self-dual Yang-Mills 
theory. The classical solution of this theory strongly resembles the 
Bethe Ansatz solution of integrable two-dimensional models. Moreover, the 
one-loop amplitudes are reminiscent of those corresponding to 
anomalous conservation laws. It is known that the 
self-dual Yang-Mills theory 
is an integrable theory and is described by very simple actions 
\cite{chalmers}.
On the other hand, integrable Lagrangeans with either anomalies 
\cite{75} or with non-vanishing amplitudes for particle 
production\cite{cosmas}
are known and are well documented in the literature. It remains an 
interesting open problem to see whether the quasi-integrability idea is the 
most efficient framework for the description of non-trivial dynamics in 
theories with higher conservation laws, in general space-time 
dimensions, in spite of the Coleman-Mandula no-go theorem 
\cite{coleman} and its more general version \cite{haag}. 

The self-dual Yang-Mills equations can be solved in two 
different
ways. The first method uses the zero curvature condition for $A_\pm$
which are solved by means of the introduction of group valued elements,
$g$ and $h$, leading to the definitions
\bear
-\frac{ie}{\sqrt{2}} A_{0+z} = g^{-1} \partial_{0+z}  \quad , \quad
-\frac{ie}{\sqrt{2}} A_{x-iy} =
g^{-1} \partial_{x-iy} g \quad ,
\nonumber \\
-\frac{ie}{\sqrt 2} A_{0-z} =
h^{-1} \partial_{0-z} h \quad, \quad
  -\frac{ie}{\sqrt 2} A_{x+iy} =
h^{-1} \partial_{x+iy}  h \quad . \nonumber
\eear
We can write the remaining information in terms of $H=gh^{1-}$, obtaining for
$H$ the equation
\be
\partial_{0-z} ( H^{-1} \partial_{0+z} H ) -
  \partial_{x+iy} ( H^{-1} \partial_{x-iy} H ) = 0 \,,
\ee
which is the Euler-Lagrange
equation of an action proposed by Donaldson, and by
Nair and Schiff\cite{don-na-sch}, and which reads
\bear
S[H] &=& \frac{f_\pi^2}{2} \int d^4x \tr\left(\partial_{0+z} H 
\partial_{0-z} H^{-1} -\partial_{x-iy} H \partial_{x+iy} H^{-1} \right)
  \nonumber\\
&& + \frac{f_\pi^2}{2} \int d^4x d t
\tr \left( \, [ H^{-1} \partial_{0+z} H, H^{-1}
    \partial_{0-z} H ] - \right.\nonumber\\
  &&  \left. [ H^{-1} \partial_{x-iy} H, H^{-1}
    \partial_{x+iy} H ]   \right)  H^{-1} \partial_t H \,.
\eear

On the other hand, if one interchanges equations of motion and
gauge prepotential definitions, using
\be
A_{x+iy} = 0 \quad , \quad A_{0+z} = \sqrt 2\partial_{x+iy} \Phi 
\quad , \quad A_{x-iy} =\sqrt 2  \; \partial_{0-z} \Phi \quad ,
\ee
one obtains the Leznov action\cite{leznov}
\be
S_{Leznov} = f_\pi^2 \int d^4x  \tr \left({1\over 2} \partial 
\phi \cdot \partial \phi +{ie\over 3}\phi \left[\partial_{x+iy} 
\phi,\partial_{0-z} \phi \right] \right)
\ee
whose equation of motion is equivalent to the remaining equation for
self-dual Yang-Mills, that is
\be
\partial^2\Phi -i e  \left[ \partial_{x+iy}\Phi,
\partial_{0-z} \Phi \right] = 0
\ee
There is no proof of equivalence of both (namely DNS and Leznov action),
but there are hints in this direction. It is the interactive solution
of Leznov equation which is of the type of series predicted by the
Bethe Ansatz approach to two-dimensional integrable models. The detailed
analysis of Cangemi\cite{cangemi} leads to that direction in a clear
way.

\chapter{Conclusions and Final Remarks}
Two-dimensional quantum field theory provides a very powerful laboratory for 
obtaining insight into the non-perturbative aspects of quantum field theory. 
The kinematical simplifications which occur in two dimensional space-time 
have allowed for the complete solution of a variety of models involving
interacting fields. The non trivial nature of these solutions provide a deeper
insight into the structure of quantum field theory, and has found useful
applications in several areas of research, such as string theories and systems
of statistical mechanics at criticality.

In all cases of completely solvable models, free bosonic fields and their
exponentials, as well as the boson-fermion equivalence for free fields play 
a key role in the explicit construction of the correlation functions. The
extension of these results to the case of massive fermions takes us, strictly
speaking, outside the realm of free fields. Nevertheless, by doing an expansion
in the mass parameter, the equivalence of the massive Thirring model to the
sine Gordon equation could be proven on the basis of zero mass bosonic fields
alone. Mandelstam's representation of fermions in terms of bosons
showed that the fermion of the massive Thirring model could be identified with
the soliton of the sine Gordon theory. The existence of such soliton sectors
provides a link with the order disorder algebra in statistical
mechanics. The existence of an infinite number of conservation
laws at the classical and quantum level,  eventually allowed 
the exact S matrix in both the soliton-soliton sector and soliton-bound-state
sector, to be constructed. These ideas prove useful in more
algebraic approaches to $QFT$.

The study of two dimensional gauge theories provides a deeper insight into 
features which are believed  to be shared by four dimensional chromodynamics. 
Among these properties are the $\theta$-vacuum, the screening of
``colour" and confinement of quarks, the
Nambu-Goldstone realization of the $\eta'$-the so called $U(1)$ problem- and
the topology of the configuration space over which one functionally integrates.

Conservation laws, $1/N$ expansion, $S$-matrix factorization and operator
techniques, were essential tools for the non perturbative treatment of the
$O(N)$ and $SU(N)$ chiral Gross-Neveu models, non linear sigma
models and $\Complex P^{N-1}$ models.
These models turned out to be  much more complex than a $U(1)$ gauge theory
of fermions interacting  with gauge fields  via  minimal coupling. In the 
case of sigma models, a very rich structure emerged. For this theory,
geometry was the main guideline, since following the ideas developed
in the context of gravity, one wished to formulate a theory in terms of the
largest possible number of symmetries. This was discussed within the framework
of differential geometry at the classical level. The outcome was a very elegant
description of integrable theories, allowing for a Lax pair at the classical
level, thus generalizing the ideas developed for the sine Gordon theory. At
the quantum level, one obtains field theories with exactly computable
$S$-matrices. Integrability properties of these sigma
models are found to hold also for a classical supersymmetric Yang-Mills theory
and supergravity in 10 dimensions, as well as for the respective theories 
obtained by dimensional reduction in 4 dimensions. This opens the possibility 
for non-trivial models in higher dimensions.

One of the realistic applications of sigma models, is in the realm of 
string theories and quantum gravity, where the background field method can be 
used to calculate the quantum corrections to the Einstein's
equations. The requirement of conformal invariance in the framework
of string theory, enforces the vanishing of the sigma model $\beta$ function,
which turns out to be the ``string corrected" gravitational equations of 
motion.

The construction of non-trivial S-matrices on the basis of factorization, and
its relation to an infinite number of conservation laws,
has useful applications in algebraic quantum field
theory. In fact, conformally invariant theories are known to possess a rich
algebraic structure derived from the fusion rules of the elementary fields.
These fusion rules being associative, and at the same time non-trivial, imply
certain algebraic relations known from the mathematical literature as Artin
Braid relations. These relations are similar to the factorization relations
obtained for factorizable S-matrices. A general structure commonly known as
``quantum groups", or to be more precise, ``quantum algebras", emerge naturally
from these constructions.

The anomalous two dimensional chiral gauge theories
provide a useful laboratory for understanding the role played by gauge
anomalies in chiral gauge theories. In this respect,
chiral $QCD_4$ shares many properties with
two dimensional QCD. In fact, in any attempt to
understand string theories away from criticality, or to deal with the
Weinberg-Salam model of the weak interactions one
must deal with the problem of chiral anomalies. 

One of the most promising applications of two-dimensional models to the
description of fundamental interactions is that of string theory. 
Strings describe a two-dimensional world-sheet embedded in the target
manifold, in which the string moves. The dynamic is thus governed by two
dimensional quantum field theory. Because of reparametrization invariance of 
string theory, conformal invariance can be extensively used to understand the
string. In fact, interaction of strings is formulated by using
vertex operators, which are generalizations of the exponential fields.

The old string model, as well as the Neveu-Schwarz-Ramond model, could be
described by a two dimensional local Lagrangian. The
discovery of space-time supersymmetry by means of the
Gliozzi-Scherk-Olive construction, permitted 
one to envisage possible applications to grand unified
theories, since supersymmetry was a requirement to solve the hierarchy problem
in unified theories; this use of string theories had been proposed earlier,
from the interpretation of the spin 2 (lowest) state of closed string theory,
as the graviton, and strings would in such a case be a theory of quantum
gravity. This unification attempt was later enhanced
by important works on the cancellation of anomalies. This led to a
new superstring theory in which left and right 
movers were treated differently, i.e.,  the heterotic string.

Of course, our ultimate objective is to understand the dynamics of the real 
world. Many successes of quantum field theory merely rely on  kinematical
arguments, such as the idea of dynamical symmetry
breaking and the construction of representations of the gauge groups in the
matter sector. The actual non-perturbative dynamics of quantum field theory in
four dimensions remains largely unknown. Hence, applications of the
experiences gained from the study of two dimensional $QFT$, to higher
dimensions is highly sought and scores success. It is rewarding that the
techniques of two dimensional $QFT$ provides many of the expected results,
sometimes just ``taken for granted" in higher dimensions, due to the lack of
methods for proving or disproving their validity. We have however also
witnessed some surprising features which are non-perturbative in nature and
are difficult to be understood in a perturbative context.

Nevertheless, some important results concerning generalizations to higher 
dimensions have already
been obtained. Bosonization of fermions is also possible in three dimensional
space-time if we have a gauge field with a Chern-Simons density in the
Lagrangian. Three dimensional models have been studied in this context. The 
ideas developed in the context of two dimensional quantum field theory thus 
appear to represent a step in the right direction. Moreover, further
recent progress made in the study of random surfaces, shows that the ideas 
of string theory have a wider range
of validity. In fact it is thought that a phase transition in string
theory occurs, such that the theory can
accommodate the grand unification ideas, statistical models, and strong
interactions, in a common framework.

More recently, direct use of the integrability ideas in order
to establish a computational procedure for the high energy limit of
non-abelian gauge theory has been advocated. The results seem promising,
and one envisages the possibility of describing the gluon interactions
in terms of a two-dimensional integrable spin system. Moreover, it is 
also possible to describe the vast complications of perturbative
amplitudes of Yang-Mills theories in terms of a single component
field described by a very simple action connected with self-dual Yang-Mills 
theory, which is by far simpler than the original theory itself.

{\bf Acknowledgements}: I would like to thank R. Mohayaee for discussions
and corrections of the manuscript, and CNPq and FAPESP for financial
support.


\end{document}